# CFD Optimization of an Orifice for Cavitation Activity

**Hrutuj Raut**

**Department of Chemical Engineering, Vishwakarma Institute of Technology, Pune, India**

## Abstract

A total of 3 orifice plates were observed – 10-, 9- and 5-hole. With each orifice plate, 4 pressure profiles were compared – 10, 7, 4 and 1 bar for each Pressure and Velocity Profiles were compared. Shnerr-Sauer Model was used in the ANSYS Fluent for CFD. An extensive comparison has been made between every model with respect to Pressure and Velocity Profiles, Bubble Radius and Cavitation Number. This has been an acute study over cavitation in a multi-hole orifice plate for which the preceding data has been observed and validated.

**List of abbreviations and symbols:** d: hole diameter; D: Plate diameter; ρ: density; $h_m$: Kinetic Energy; $h_t$: Thermal Energy; $h_c$: Chemical Energy; φ: Potential Energy; R: Mass Transfer Rate; $R_B$: Bubble Radius; p: pressure; v: flow velocity; We: Weber number; l: characteristic length; σ: surface tension; $p_r$: recovered pressure; $p_v$: vapor pressure of the fluid; CFD: Computational Fluid Dynamics; CN: Cavitation Number

# Table of Contents

# List of Figures

# List of Tables

# List of Equations

# Introduction

## Background

An orifice metre consists of a flat plate with a precisely machined, sharp-edged hole concentrically positioned in a pipe as shown in Figure 1. As liquid passes through the pipe, the flow abruptly decreases as it gets closer to the orifice and then abruptly increases as it reaches the orifice and returns to the full pipe diameter. This creates a throat or *vena contracta* just after the orifice. Similar to a venturi metre, this change in the *vena contracta's* flow pattern promotes higher velocity and, as a result, decreased neck pressure. (Brennen, 1995)

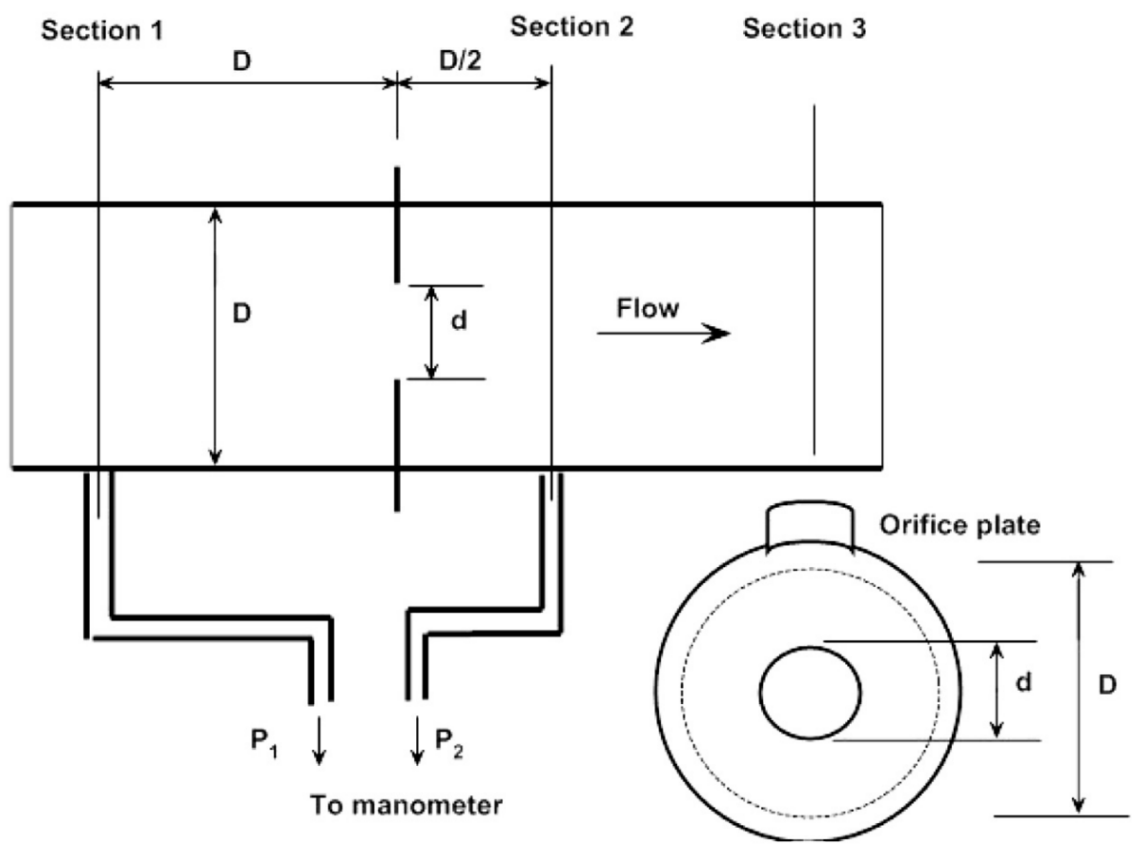

Figure 1: Orifice meter (2022)

Using formulae created previously for the venturi metre and the flow nozzle, the pressure differential between section 1, with the entire flow, and section 2, at the throat, may then be utilised to estimate the liquid flow rate. The orifice meter's coefficient of discharge C is substantially lower than that of a venturi metre or a flow nozzle because of the abrupt contraction at the orifice and the abrupt expansion that occurs following it. (Brennen, 1995)

For different values of the pipe Reynolds number, C is shown to vary with the beta ratio d/D in Figure 2.

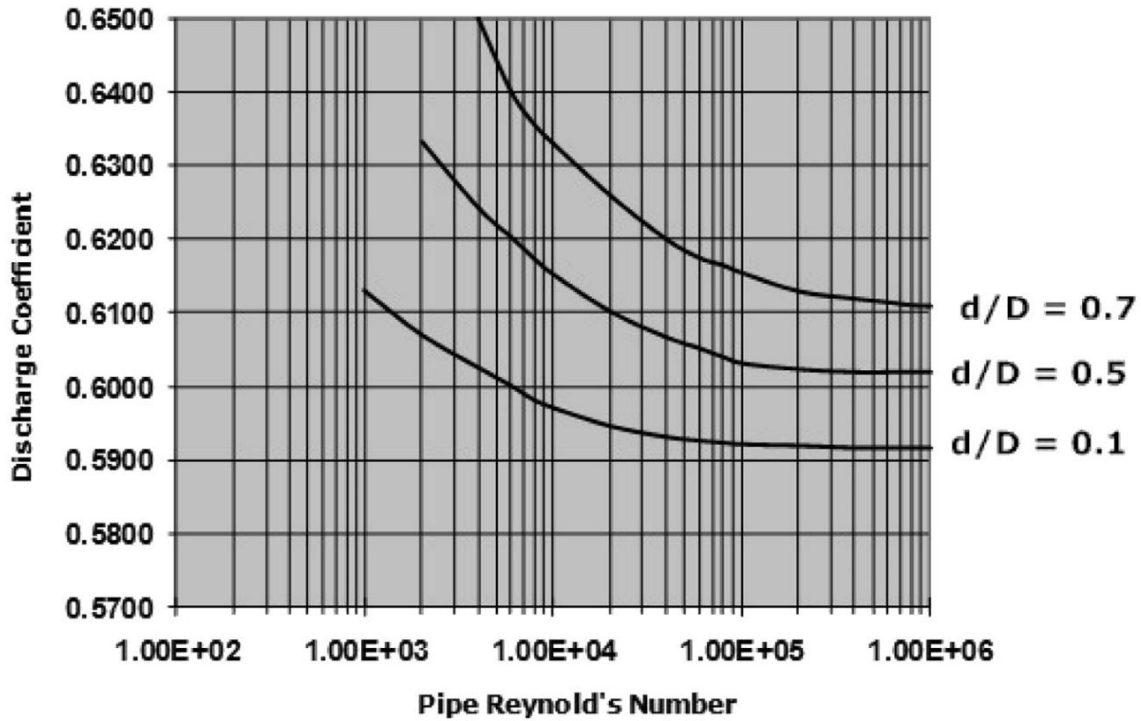

Figure 2: Orifice Meter Discharge Coefficient (2022)

The orifice plate has the most energy loss as a result of the abrupt constriction and subsequent abrupt expansion. However, due to the smooth, gradual reduction at the throat and the smooth, gradual expansion after the throat, the venturi metre has a lower energy loss when compared to a flow nozzle. (Brennen, 1995)

When a fluid's pressure falls below its vapour pressure, a phenomenon known as cavitation takes place in the fluid that results in the formation and subsequent collapse of vapour bubbles. Cavitation may happen in a range of applications, such as pumps, propellers, valves, and hydraulic systems, and it can happen in any fluid, including water, oil, or air. A fluid begins to boil and produce vapour bubbles when its pressure falls below that of its vapour pressure. (Dabiri, et al., 2007)These bubbles can range in size from extremely small (micrometres or less) to very large (centimetres or larger). The vapour bubbles rupture as the fluid's pressure rises once more, generating shock waves that might harm the nearby materials. (Brennen, 1995)

High temperatures and pressures may be produced by the collapse of the vapour bubbles, which may cause erosion and pitting of the material. Additionally, the burst of bubbles may cause noise and vibrations that could interfere with the smooth operation of machinery and equipment. By increasing the fluid's pressure, decreasing its velocity, or raising its viscosity, cavitation can be avoided. To lower the danger of cavitation, it may occasionally be required to adjust the design of the machinery or equipment. In conclusion, cavitation is a complicated phenomena that can harm machinery and equipment. It can be avoided by taking the proper steps to raise the fluid's pressure, lower its velocity, or increase its viscosity. (Ebrahimi, et al., 2017)

# Governing Equations

This chapter explains the governing equations solved with FLUENT and the turbulence and cavitation models used for this simulation. Two equation models are used for the simulations. The flow equations and energy equations are also described in detail.

## Flow calculations

The flow is governed by the continuity equation, the energy equation, and the Navier-Stokes momentum equations. Mass, energy and momentum are transferred by convective flow and diffusion of molecules and turbulent eddies. (S.Shah, et al., 2011)

## Continuity Equation

This equation describes the conservation of mass and is written as,

$$\frac{\partial \rho}{\partial t} + \frac{\partial \rho U_1}{\partial x_1} + \frac{\partial \rho U_2}{\partial x_2} + \frac{\partial \rho U_3}{\partial x_3} = 0$$

**Equation 1: Continuity Equation**

Defines the rate of increase of mass in a control volume as an equal amount over its area.

## Navier-Stokes Equation

Momentum balance, also known as the Navier-Stokes equations, obeys Newton's second law: The change in momentum in all directions is equal to the sum of the forces acting in those directions. Two different kinds of forces act on a finite volume element, surface forces and body forces. Surface forces include pressure and viscous forces, and body forces include gravitational, centrifugal, and electromagnetic forces. (S.Shah, et al., 2011)

This equation can be written as (for a Newtonian fluid),

$$\frac{\partial U_i}{\partial t} + U_j \frac{\partial U_i}{\partial x_j} = -\frac{1}{\rho}\frac{\partial \rho}{\partial x_i} + \nu \frac{\partial}{\partial x_j}\left(\frac{\partial U_i}{\partial x_j} + \frac{\partial U_j}{\partial x_i}\right) + g_i$$

**Equation 2: Navier-Stokes Equation**

In addition to gravity, there may be other external sources that can affect the acceleration of the fluid, such as electric and magnetic fields. Strictly speaking, it is the momentum equations that make up the Navier-Stokes equations, but sometimes the continuity and momentum equations together are called the Navier-Stokes equations. The Navier-Stokes equations are limited to macroscopic conditions. (Bashir, et al., 2011)

The continuity equation is difficult to solve numerically. In CFD programs, the continuity equation is often combined with the momentum equation to form Poisson's equation. For constant density and viscosity, the equation can be written as,

$$\frac{\partial}{\partial x_i}\left(\frac{\partial P}{\partial x_i}\right) = -\frac{\partial}{\partial x_i}\left(\frac{\partial (\rho U_i U_j)}{\partial x_j}\right)$$

**Equation 3: Poisson's Equation**

This equation has more suitable numerical properties and can be solved by suitable iterative methods.

## Energy Equation

Energy is present in a flow in many forms, i.e., as kinetic energy due to the mass and velocity of the fluid, as thermal energy and as chemically bound energy. So, the total energy can be defined as the sum of all these energies. (Bashir, et al., 2011)

$$h = h_m + h_T + h_C + \Phi$$

**Equation 4: Energy Equation**

| | |
|---|---|
| $h_m = \frac{1}{2}\rho U_i U_i$ | Kinetic Energy |
| $h_T = \sum_n m_n \int_{T_{ref}}^{T} C_{p,n} dT$ | Thermal Energy |
| $h_C = \sum_n m_n h_n$ | Chemical Energy |
| $\Phi = g_i x_i$ | Potential Energy |

**Table 1: General energy equations**

In the above equations, $m_n$ and $C_{p,n}$ are the mass fraction and specific heat for species n. The transport equation for the total energy can be written using the above equations. The coupling between the energy and momentum equations is very weak for incompressible flows, so the kinetic and thermal energy equations can be written separately. Chemical energy is not included as no mode of transport was involved in this project. (Bashir, et al., 2011)

## Schnerr - Sauer Cavitation Model

As in the Singhal et al. model, Schnerr and Sauer follow a similar approach to derive the exact expression for the net mass transfer from liquid to vapor. The equation for the vapor volume fraction has the general form: (Babu, et al., 2018)

$$\frac{\partial}{\partial t}(\alpha \rho_v) + \nabla \cdot (\alpha \rho_v V) = \frac{\rho_v \rho_l}{\rho}\frac{D\alpha}{Dt}$$

Here, the net mass source term is as follows:

$$R = \frac{\rho_v \rho_l}{\rho} \frac{d\alpha}{dt}$$

Following a similar approach to Singhal et al., they derived the following equation:

$$R = \frac{\rho_v \rho_l}{\rho} \alpha(1-\alpha) \frac{3}{R_B} \sqrt{\frac{2}{3} \frac{(P_v - P)}{\rho_l}}$$

$$R_B = \left(\frac{\alpha}{1-\alpha} \frac{3}{4\pi} \frac{1}{n}\right)^{\frac{1}{3}}$$

R – Mass Transfer Rate

$R_B$ – Bubble Radius

## Bernoulli' Equation

Assuming a horizontal flow (neglecting the minor elevation difference between the measuring points) the Bernoulli Equation can be modified to:

$$p_1 + 1/2\, \rho\, v_1^2 = p_2 + 1/2\, \rho\, v_2^2$$

**Equation 5: Bernoulli' Equation**

Where,

p = pressure (Pa, psf (lb/ft2))

ρ = density (kg/m3, slugs/ft3)

v = flow velocity (m/s, ft/s)

## Two – Equation Models

Different turbulence models can be classified based on the number of additional equations used to close the set of equations. There are zero-, one-, and two-equation models commonly used to model turbulence. The null equation model makes a simple assumption of constant viscosity (Prandtl mixing length model). While one equation model assumes that the viscosity is related to the historical effects of turbulence in relation to the time-averaged kinetic energy. Similarly, a two-equation model uses two equations to close a set of equations. These two equations can model turbulent velocity or turbulent length scales. There are many variables that can be modelled, such as vorticity scale, frequency scale, time scale, and dissipation rate. Of these variables, the most commonly used variable is the ε dissipation rate. This model is named with respect to the modelled variables. For example, the k-ε model because it models k (turbulent kinetic energy) and k-ε (turbulent energy dissipation rate). Another important turbulence model is the k-ω model. It models k (Turbulent Kinetic Energy) and ω (Specific Dissipation Rate). These models have now become common in industrial use. These provide a significant amount of confidence because they use two variables to close the set of equations. (Jithish, et al., 2015)

## k-epsilon Model

k-epsilon Turbulence model, also known as the k-epsilon model was proposed for turbulent bounded flows. The model is divided into two equations. The first equation gives the first transported variable kinetic energy k. The equation is given as, (Yan, et al., 1990)

$$\frac{\partial(\rho k)}{\partial t} + \frac{\partial(\rho k u_i)}{\partial x_i} = \frac{\partial}{\partial x_j}\left[\frac{\mu_t}{\sigma_k}\frac{\partial k}{\partial x_j}\right] + 2\mu_t E_{ij}E_{ij} - \rho\varepsilon$$

**Equation 6: K-epsilon model**

The second equation gives the rate of dissipation of turbulent kinetic energy ε, which is the second transported variable. The equation is given as, (Yan, et al., 1990)

$$\frac{\partial(\rho\varepsilon)}{\partial t} + \frac{\partial(\rho\varepsilon u_i)}{\partial x_i} = \frac{\partial}{\partial x_j}\left[\frac{\mu_t}{\sigma_\varepsilon}\frac{\partial \varepsilon}{\partial x_j}\right] + C_{1\varepsilon}\frac{\varepsilon}{k}2\mu_t E_{ij}E_{ij} - C_{2\varepsilon}\rho\frac{\varepsilon^2}{k}$$

**Equation 7: K-epsilon model**

## k-omega Model

A two-equation model for the prediction of turbulence kinetic energy k and specific rate of dissipation ω. The two equations are given as, (Yan, et al., 1990)

$$\frac{\partial(\rho k)}{\partial t} + \frac{\partial(\rho u_j k)}{\partial x_j} = \rho P - \beta^*\rho\omega k + \frac{\partial}{\partial x_j}\left[\left(\mu + \sigma_k\frac{\rho k}{\omega}\right)\frac{\partial k}{\partial x_j}\right], \quad with\ P = \tau_{ij}\frac{\partial u_i}{\partial x_j}$$

**Equation 8: K-omega model**

$$\frac{\partial(\rho\omega)}{\partial t} + \frac{\partial(\rho u_j \omega)}{\partial x_j} = \frac{\alpha\omega}{k}\rho P - \beta\rho\omega^2 + \frac{\partial}{\partial x_j}\left[\left(\mu + \sigma_\omega\frac{\rho k}{\omega}\right)\frac{\partial \omega}{\partial x_j}\right] + \frac{\rho\sigma_d}{\omega}\frac{\partial k}{\partial x_j}\frac{\partial \omega}{\partial x_j}$$

**Equation 9: K-omega model**

The k-omega model is specifically used for the prediction of boundary-layer transition and flows with low Reynolds number.

SST models stand for Shear Stress Transport model, it is a combination of models k-omega and k-epsilon. Since k-epsilon is a high Reynolds number model, model k-omega is used in the near wall region. While in the area away from the walls, the k-epsilon model is used. The SST model uses a blending function whose value depends on the distance from the walls. At the wall, in the viscous sublayer, this blending function is used only as the k-omega model. Areas away from the wall have this function zero and only use the k-epsilon model. This model also includes a cross-diffusion term. In this model, the turbulent viscosity is modified to include the effect of turbulent shear stress transfer. Modelling constants are also different from other models. These properties make the SST model reliable for adverse pressure gradient flow and boundary layer separation. (S.Shah, et al., 2011)

# CFD Analysis

CFD analysis has been done for a 10-, 9- and 5-hole orifice plate with pressure conditions satisfying each of the following hole characteristics with 1, 4, 7 and 10 bar. Following is the CFD analysis of all considered models.

| *Geometry* | |
|---|---|
| *Length X* | 14.5 mm |
| *Length Y* | 80. mm |
| *Length Z* | 14.5 mm |
| *Inner Diameter* | 12.5 mm |
| *Outer Diameter* | 14.5 mm |
| *Thickness* | 2 mm |
| *Hole Diameter* | 1.5 mm |

**Table 2: Base Geometrical Statistics**

| *Mesh Statistics* | |
|---|---|
| *Physics Preference* | CFD |
| *Solver Preference* | Fluent |
| *Element Order* | Linear |
| *Element Size* | Default (4.1293 mm) |
| *Sizing* | |
| *Use Adaptive Sizing* | No |
| *Growth Rate* | Default (1.2) |
| *Max Size* | Default (8.2586 mm) |
| *Mesh Defeaturing* | Yes |
| *Defeature Size* | Default (2.0647e-002 mm) |
| *Capture Curvature* | Yes |
| *Curvature Min Size* | Default (4.1293e-002 mm) |
| *Curvature Normal Angle* | Default (18.0°) |
| *Capture Proximity* | No |
| *Bounding Box Diagonal* | 82.586 mm |
| *Average Surface Area* | 324.04 mm² |
| *Minimum Edge Length* | 4.7124 mm |
| *Quality* | |
| *Check Mesh Quality* | Yes, Errors |
| *Target Skewness* | Default (0.900000) |
| *Smoothing* | Medium |
| *Inflation* | |
| *Use Automatic Inflation* | None |
| *Inflation Option* | Smooth Transition |
| *Transition Ratio* | 0.272 |
| *Maximum Layers* | 5 |
| *Growth Rate* | 1.2 |
| *Advanced* | |
| *Rigid Body Behavior* | Dimensionally Reduced |
| *Triangle Surface Mesher* | Program Controlled |
| *Topology Checking* | Yes |
| *Pinch Tolerance* | Default (3.7164e-002 mm) |
| *Statistics* | |
| *Nodes* | 738069 |
| *Elements* | 3754064 |

**Table 3: Mesh Statistics**

To generate the Bubble Radius, a MATLAB Simulink chart was made. It solves the following Equation 10: Bubble Radius and Equation 11: Weber Number.

The value of Bubble Radius is calculated using Weber Number (We) as:

$$R = \frac{0.061 We \sigma}{2\rho_{liq} u_{rel}^2}$$

**Equation 10: Bubble Radius**

$$We = \rho v^2 l / \sigma$$

**Equation 11: Weber Number**

Where,

We = Weber number (dimensionless)

$\rho$ = density of fluid (kg/m3, lb/ft3)

v = velocity of fluid (m/s, ft/s)

l = characteristic length (m, ft)

$\sigma$ = surface tension (N/m)

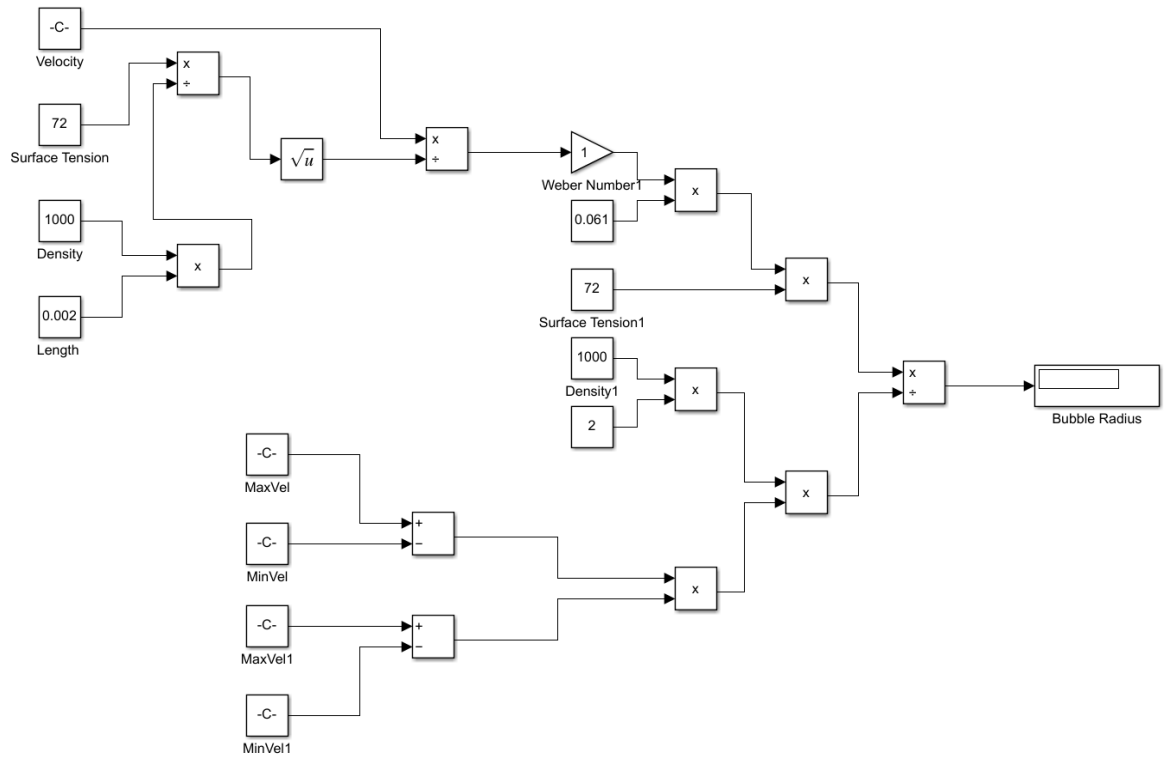

**Figure 3: MATLAB Simulink - Bubble Radius**

To generate Cavitation Number, a similar MATLAB Simulink chart was made. The Cavitation Number can be expressed as,

$$\sigma = (p_r - p_v) / (1/2\, \rho\, v^2)$$

**Equation 12: Cavitation Number**

Where,

σ = Cavitation number

$p_r$ = recovered pressure (Pa)

$p_v$ = vapor pressure of the fluid (Pa)

ρ = density of the fluid (kg/m3)

v = velocity of fluid (m/s)

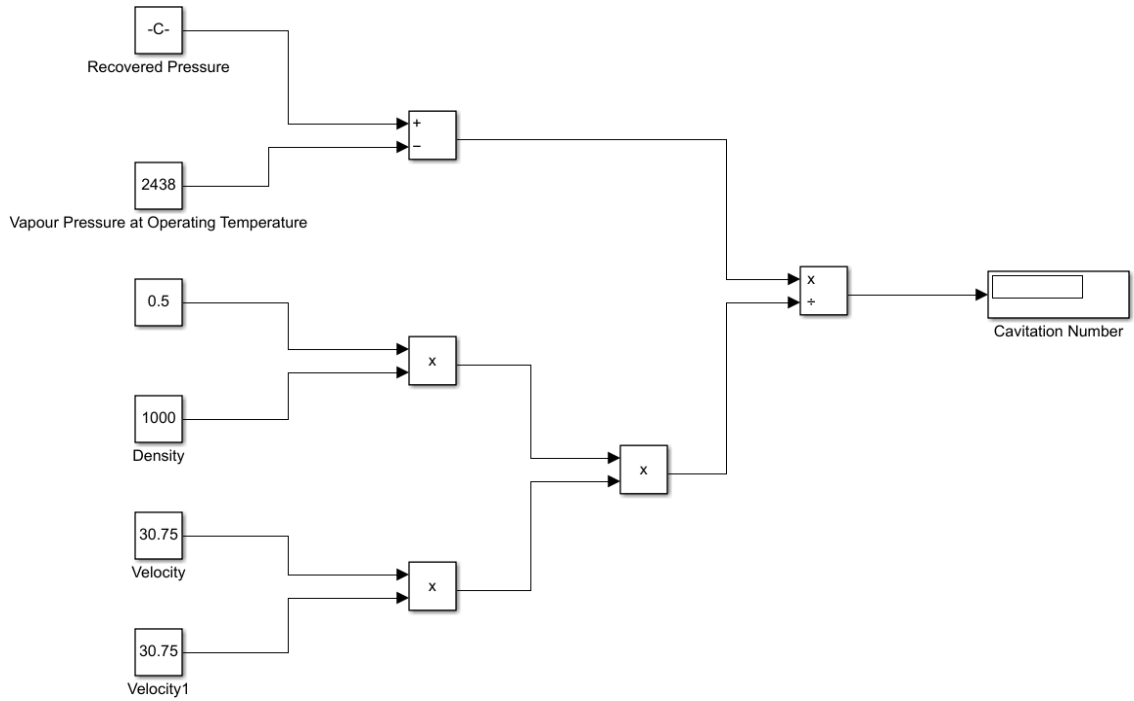

**Figure 4: MATLAB Simulink - Cavitation Number**

The recovered pressure can be calculated by Bernoulli's Equation.

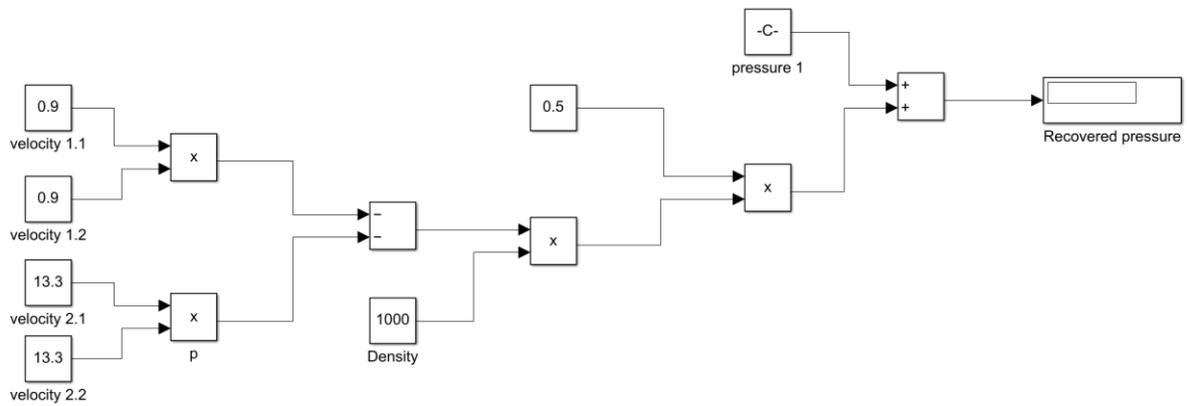

**Figure 5: MATLAB Simulink - Bernoulli's Equation**

# 10 Hole

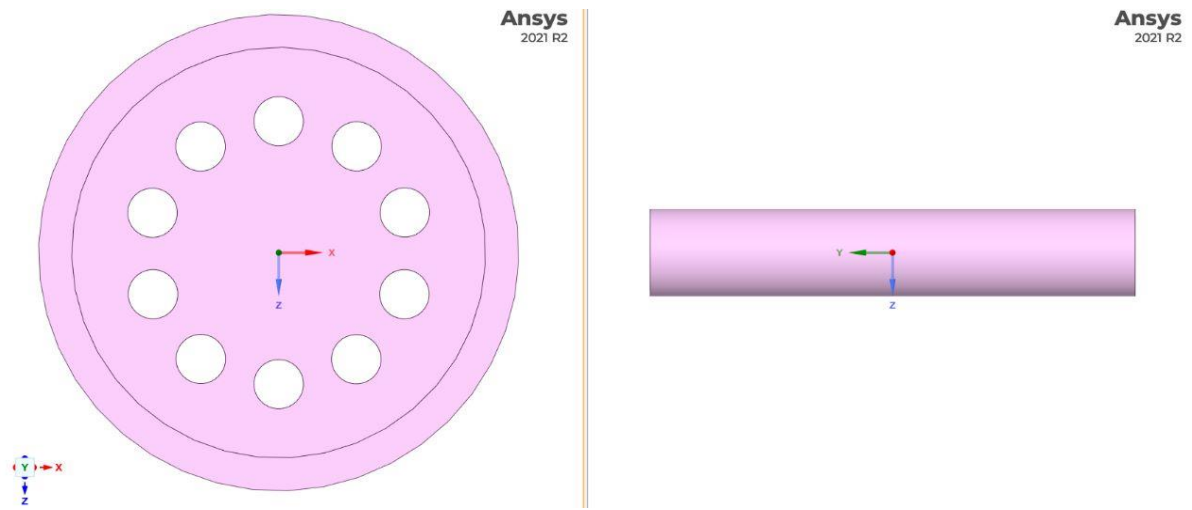

**Figure 6: 10 Hole Orifice Geometry**

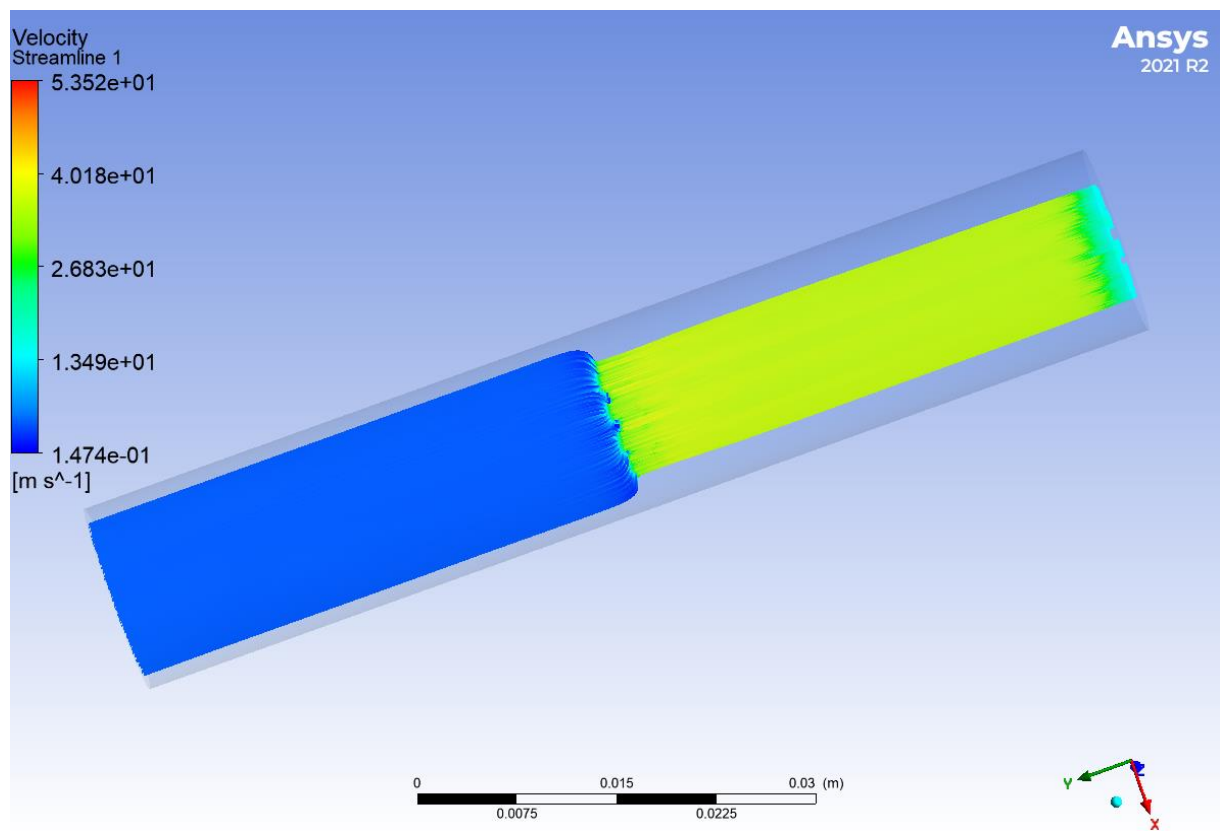

**Figure 7: Streamlines - 10 Hole**

# 10 bar

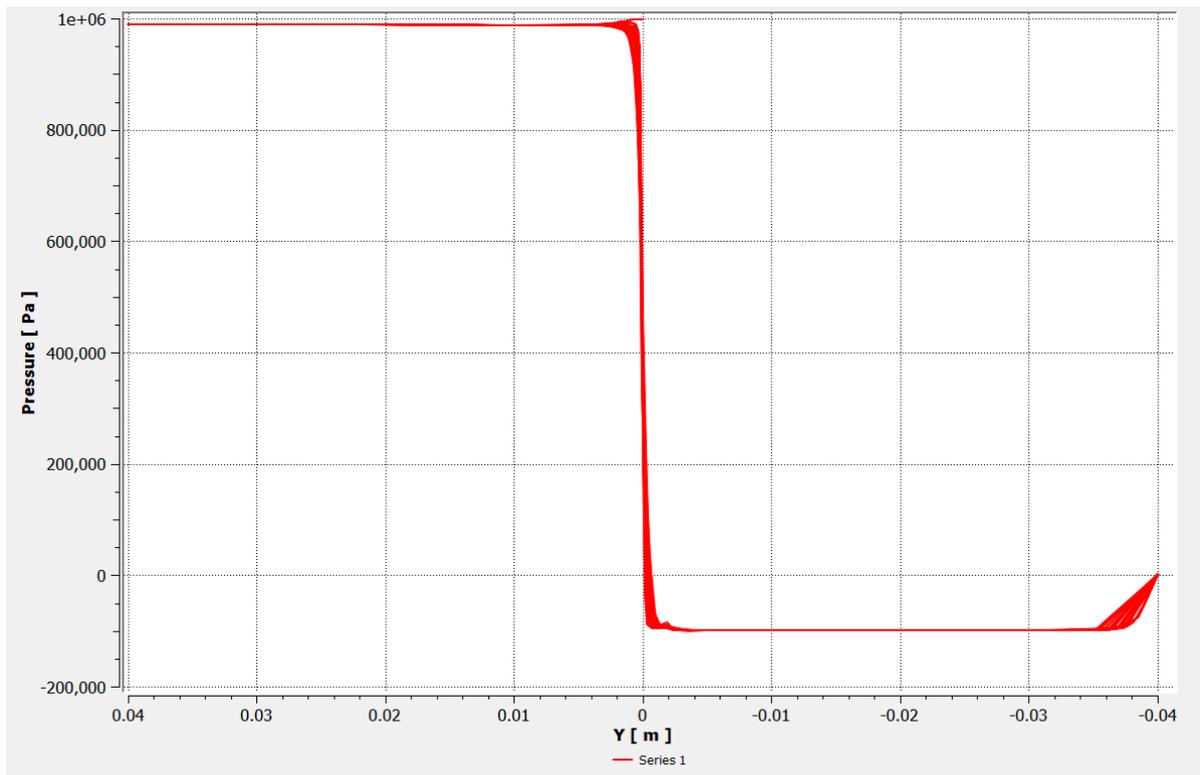

**Figure 8: Pressure Profile – 10 bar**

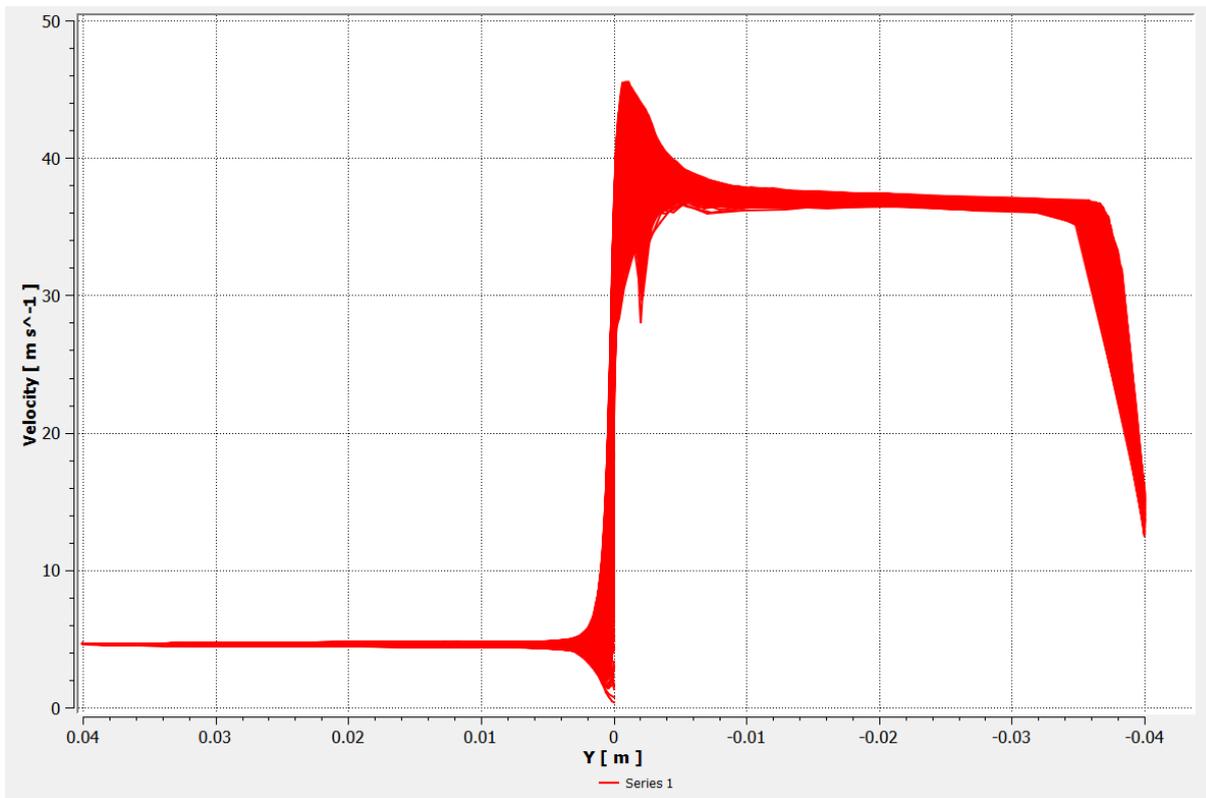

**Figure 9: Velocity Profile – 10 bar**

# 7 bar

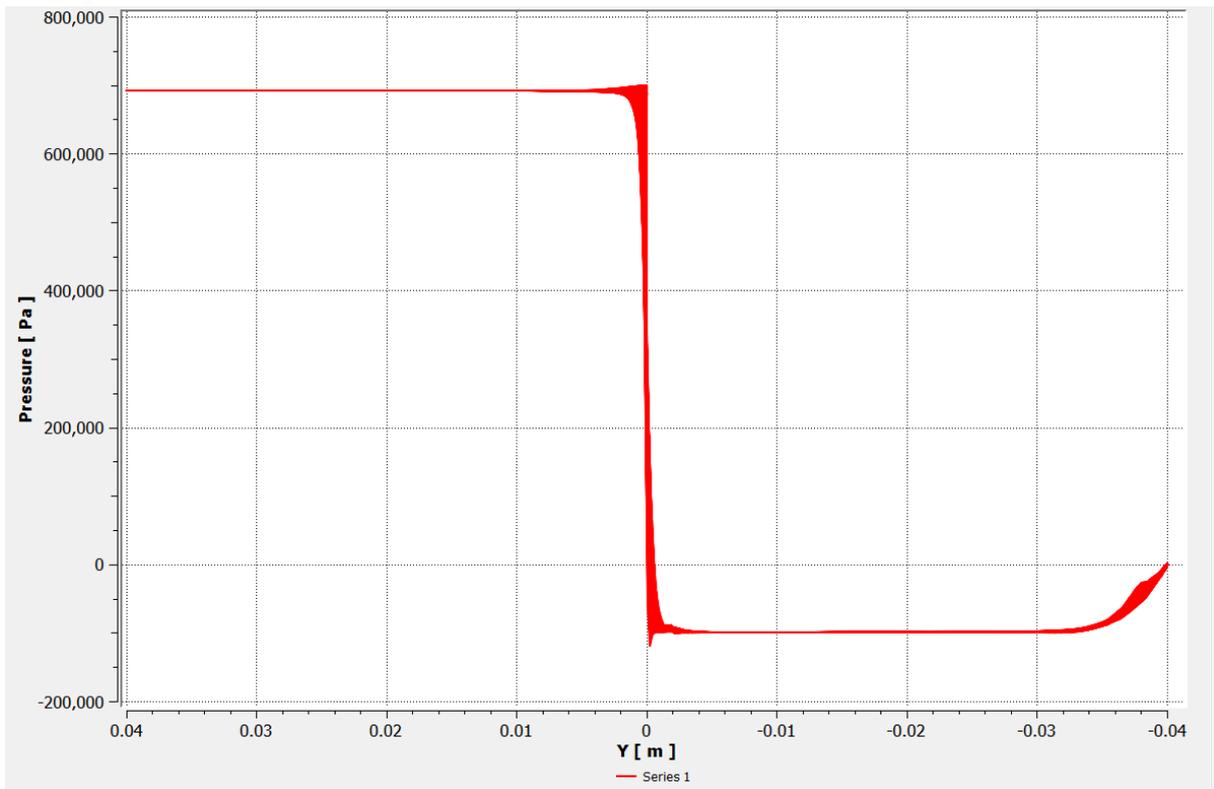

**Figure 10: Pressure Profile - 7 bar**

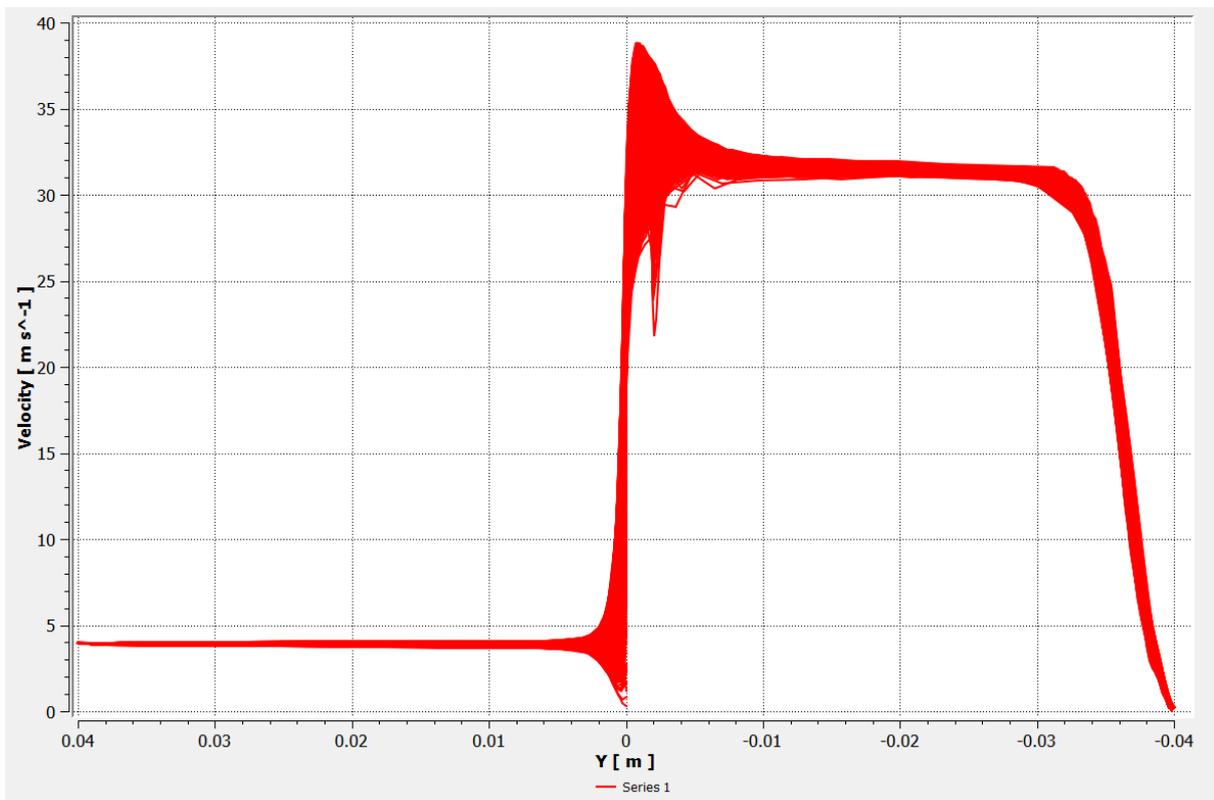

**Figure 11: Velocity Profile - 7 bar**

# 4 bar

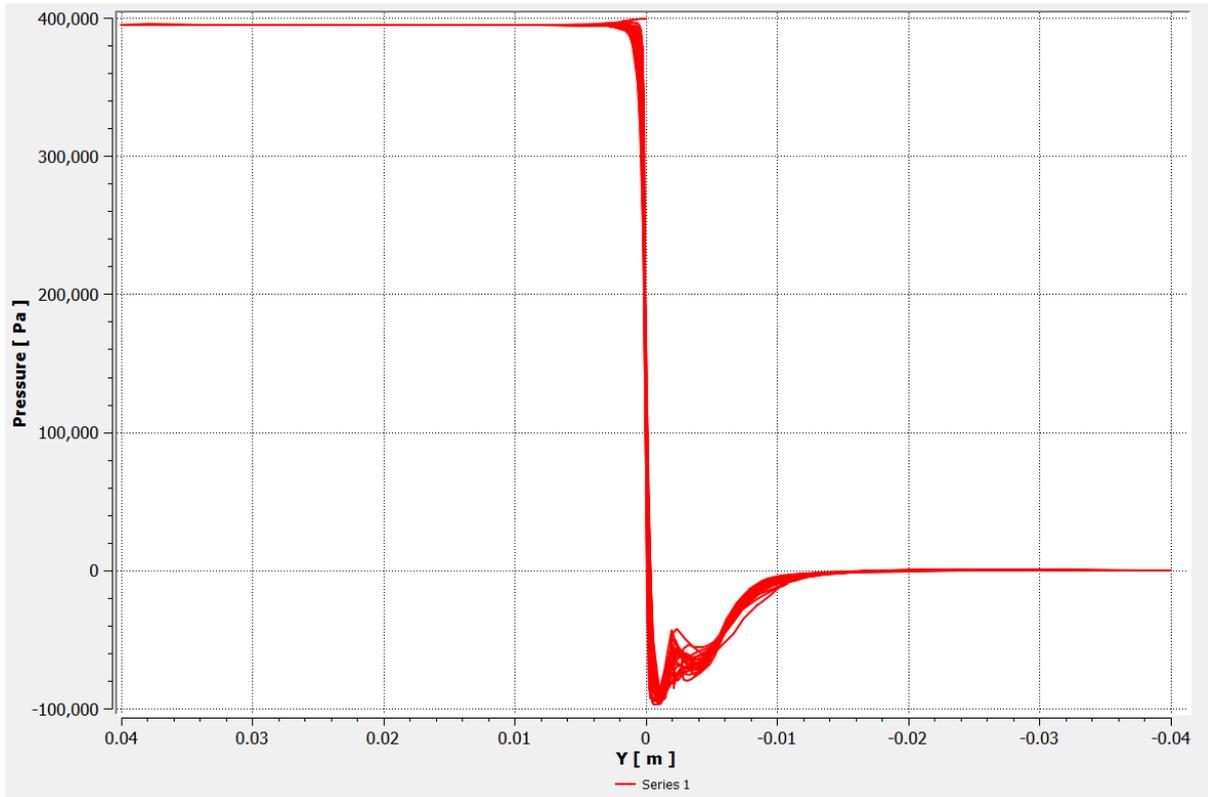

**Figure 12: Pressure Profile - 4 bar**

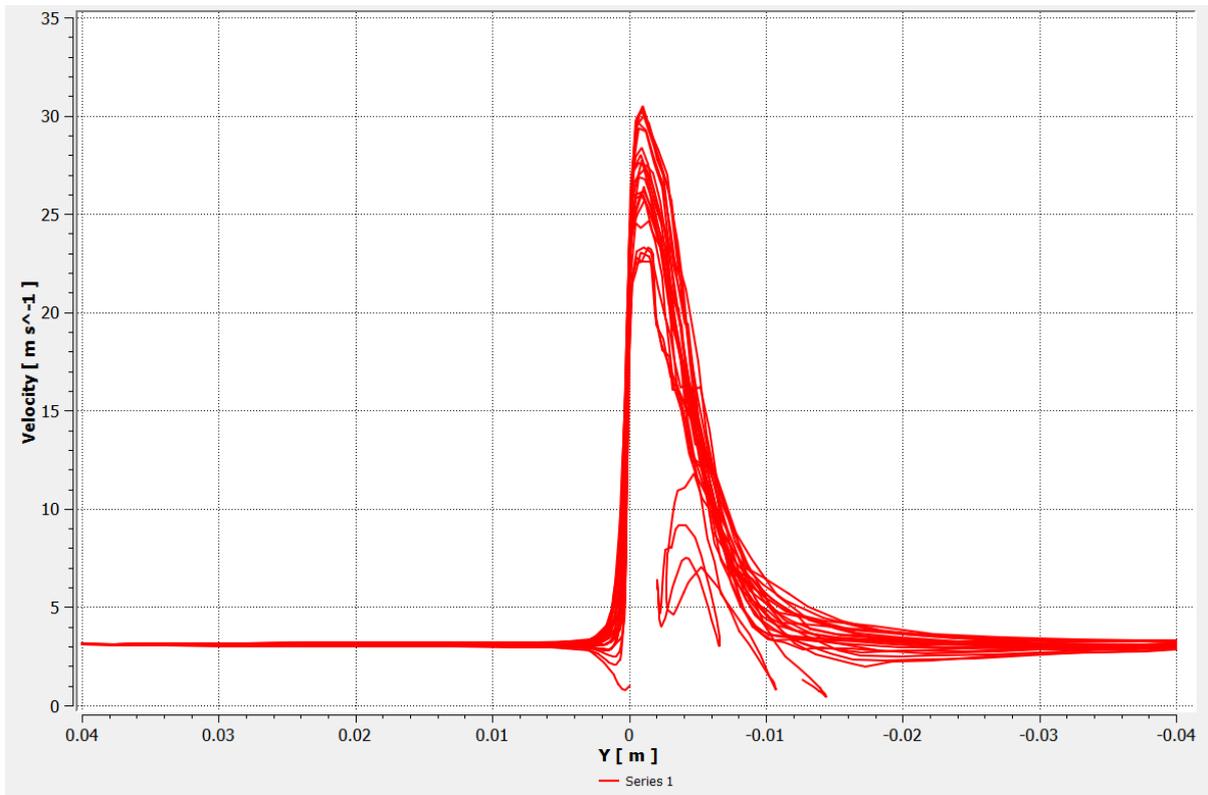

**Figure 13: Velocity Profile - 4 bar**

# 1 bar

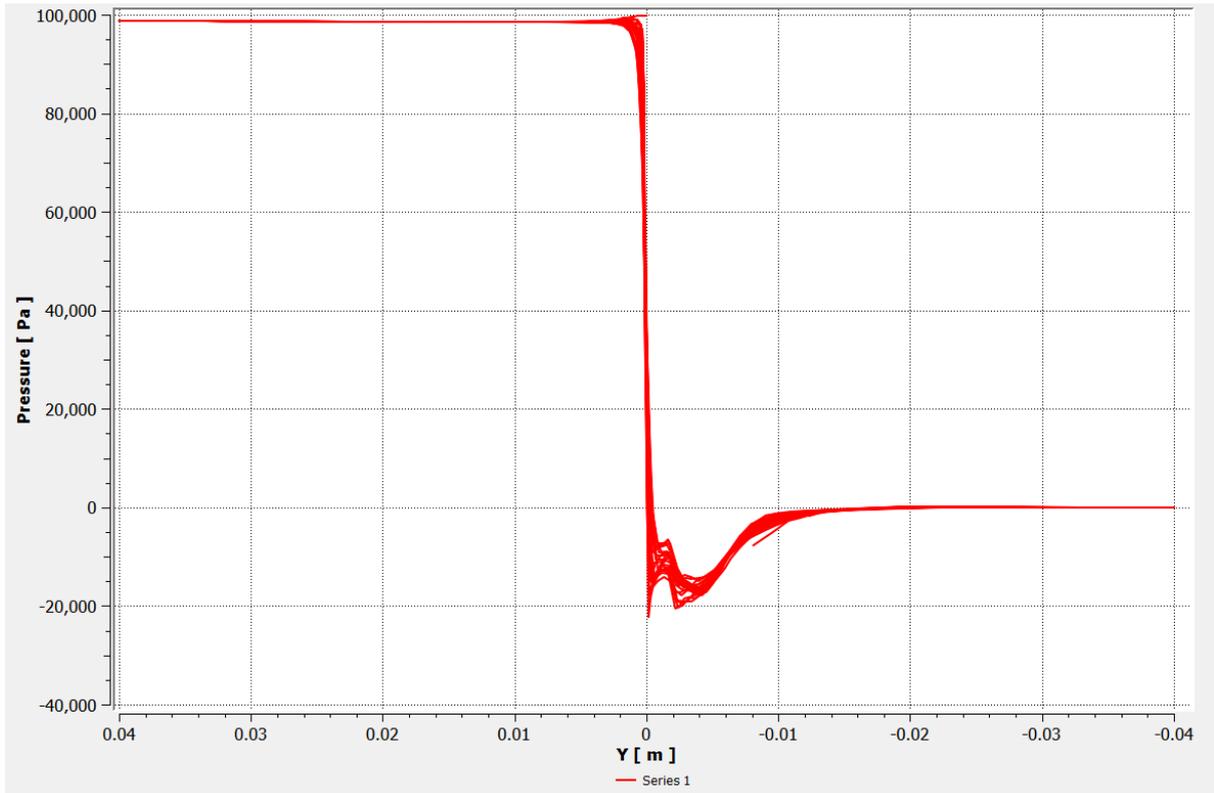

**Figure 14: Pressure Profile - 1 bar**

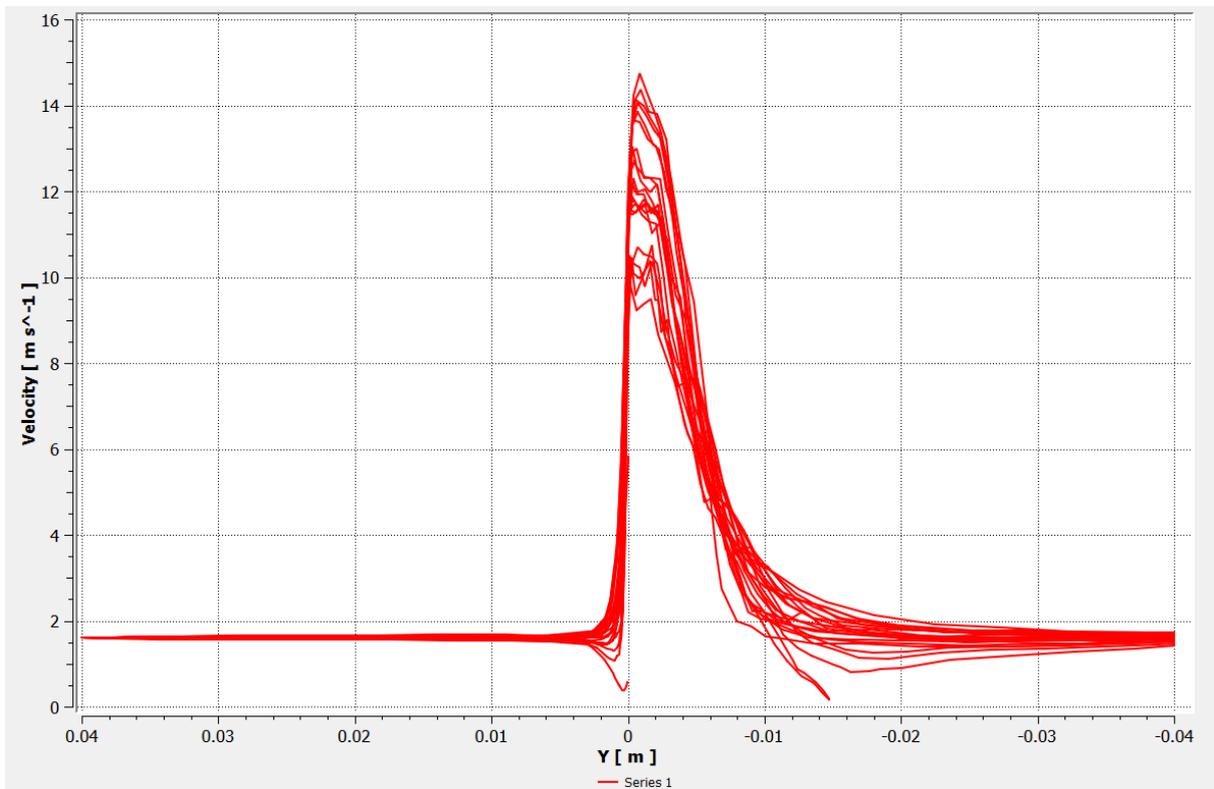

**Figure 15: Velocity Profile - 1 bar**

# Bubble Radius

*Pressure (bar)    Bubble Radius (m)*

| | |
|---:|---|
| 1 | 2.49E-05 |
| 4 | 1.20E-05 |
| 7 | 8.02E-06 |
| 10 | 6.81E-06 |

Table 4: Calculated Bubble Radius for 10 Hole Orifice Plate

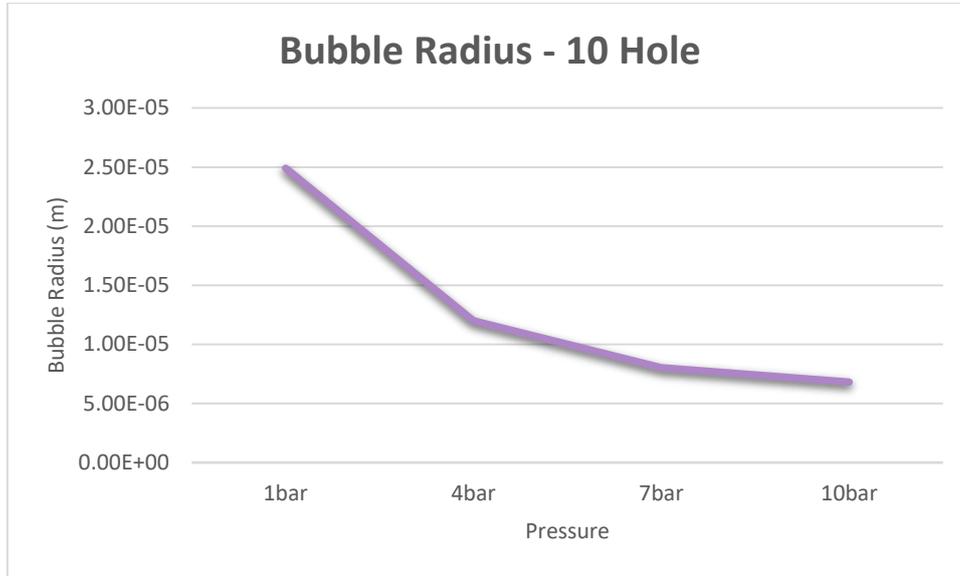

Figure 16: Graphical Representation - Bubble Radius (10 Hole)

# Cavitation Number

| Pressure (bar) | Cavitation Number |
|---:|---|
| 1 | 0.7748 |
| 4 | 0.7595 |
| 7 | 0.1 |
| 10 | 0.1329 |

Table 5: CN - 10 Hole

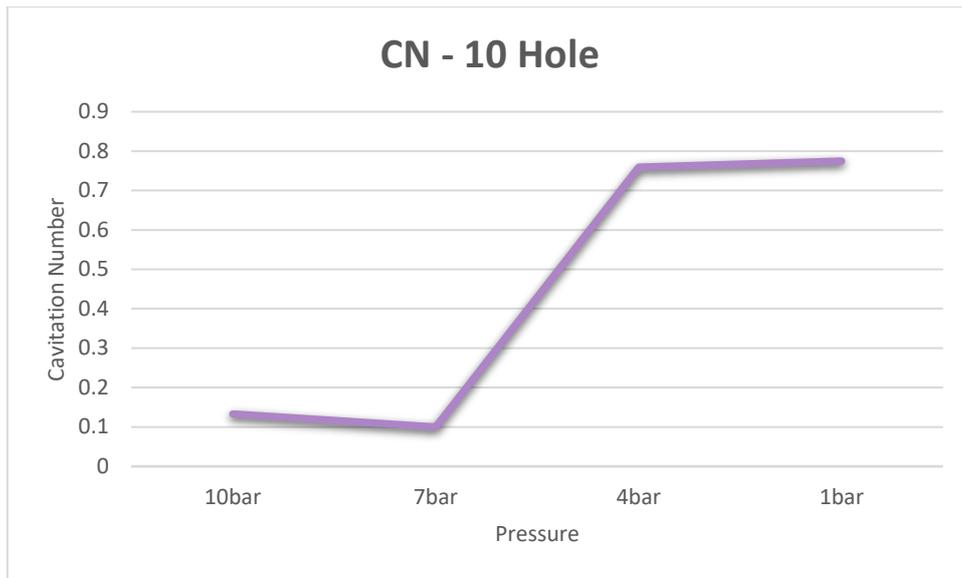

Figure 17: CN - 10 Hole

# 9 Hole

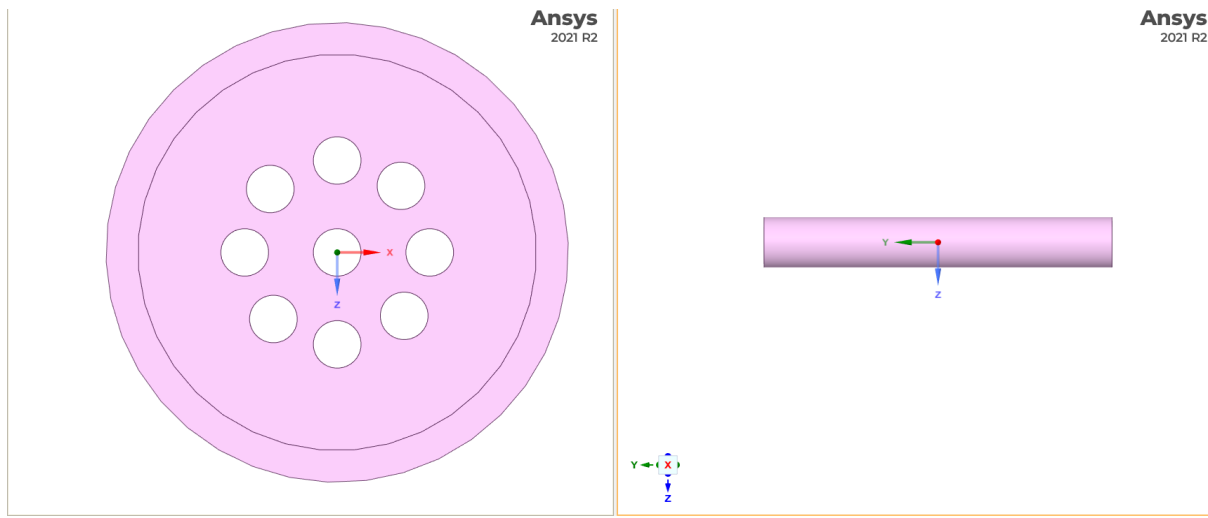

Figure 18: 9 Hole Orifice Geometry

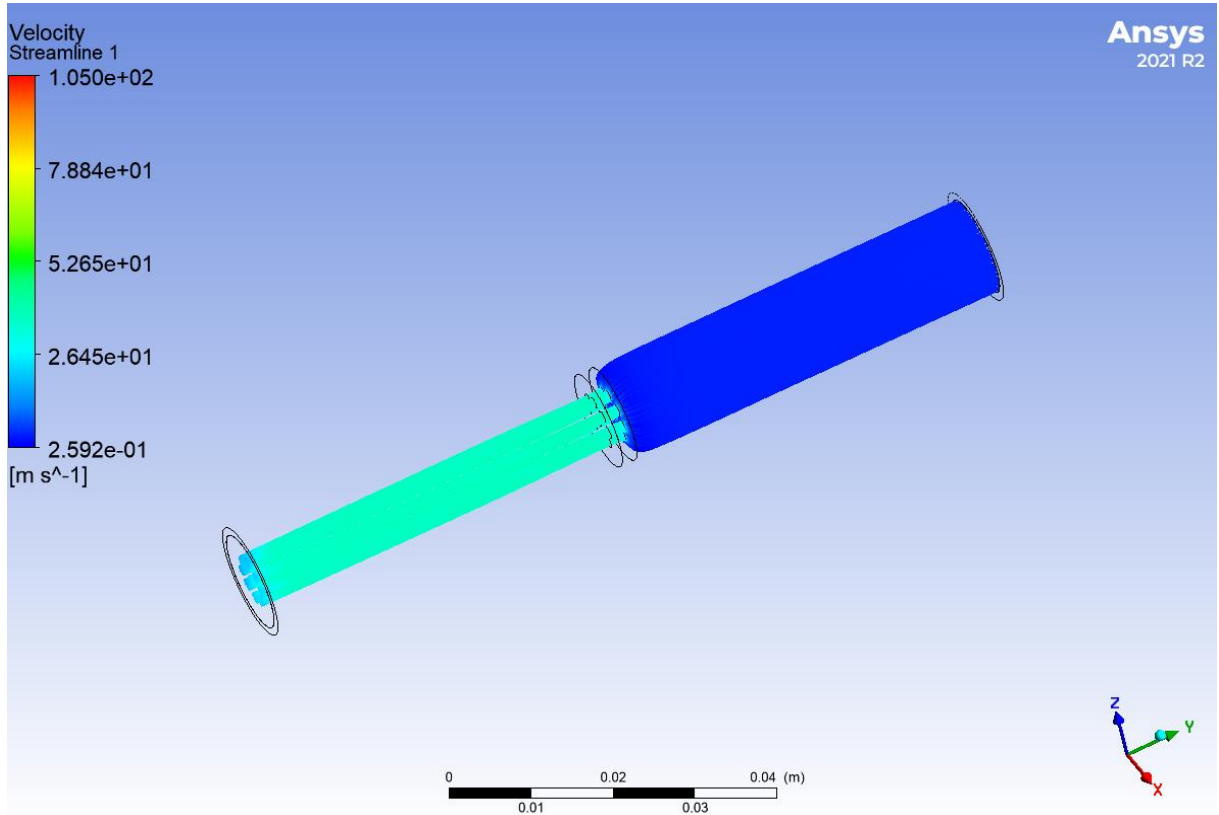

**Figure 19: Velocity Streamline - 9 Hole**

Following are the charts for different pressure and velocity profiles for each pressure condition to observe the behaviour with respect to orifice and tube.

# 10 bar

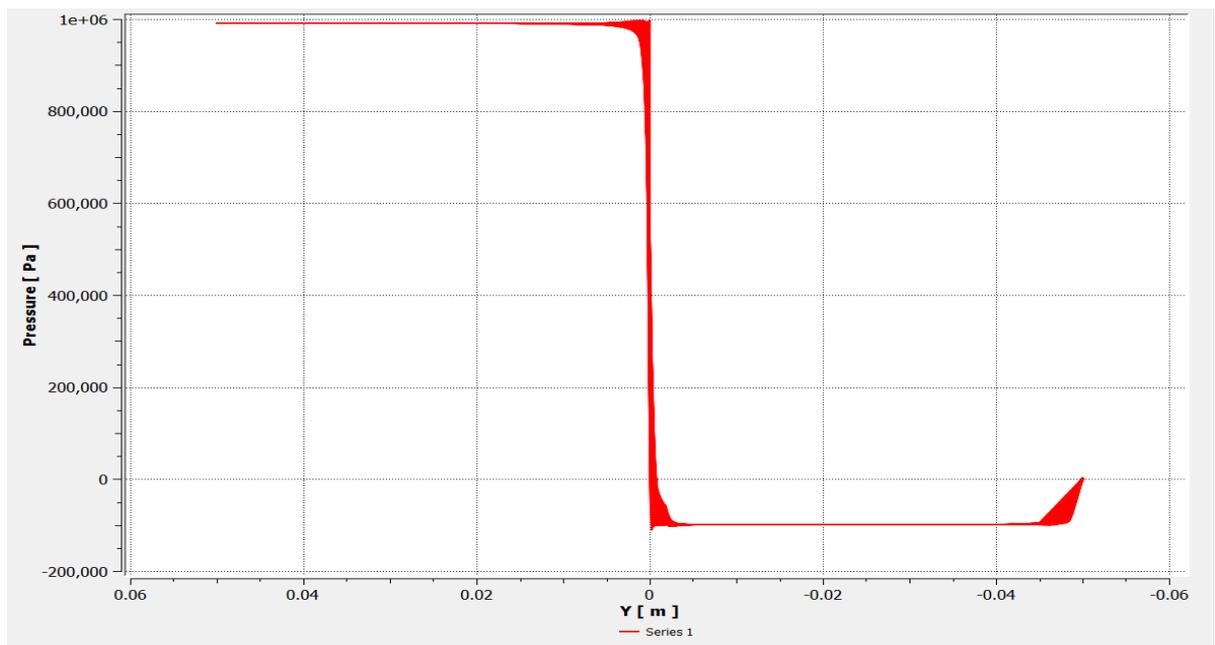

**Figure 20: Pressure Profile - 10 bar**

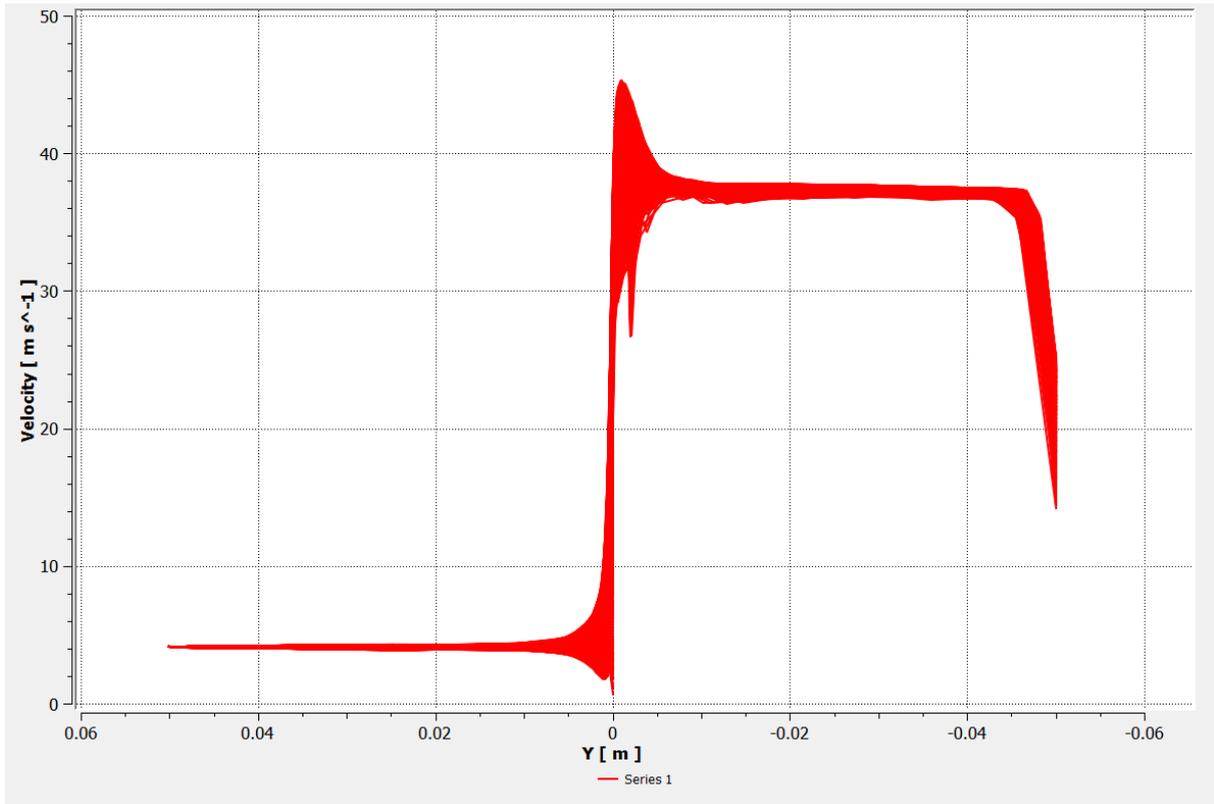

**Figure 21: Velocity Profile - 10 bar**

# 7 bar

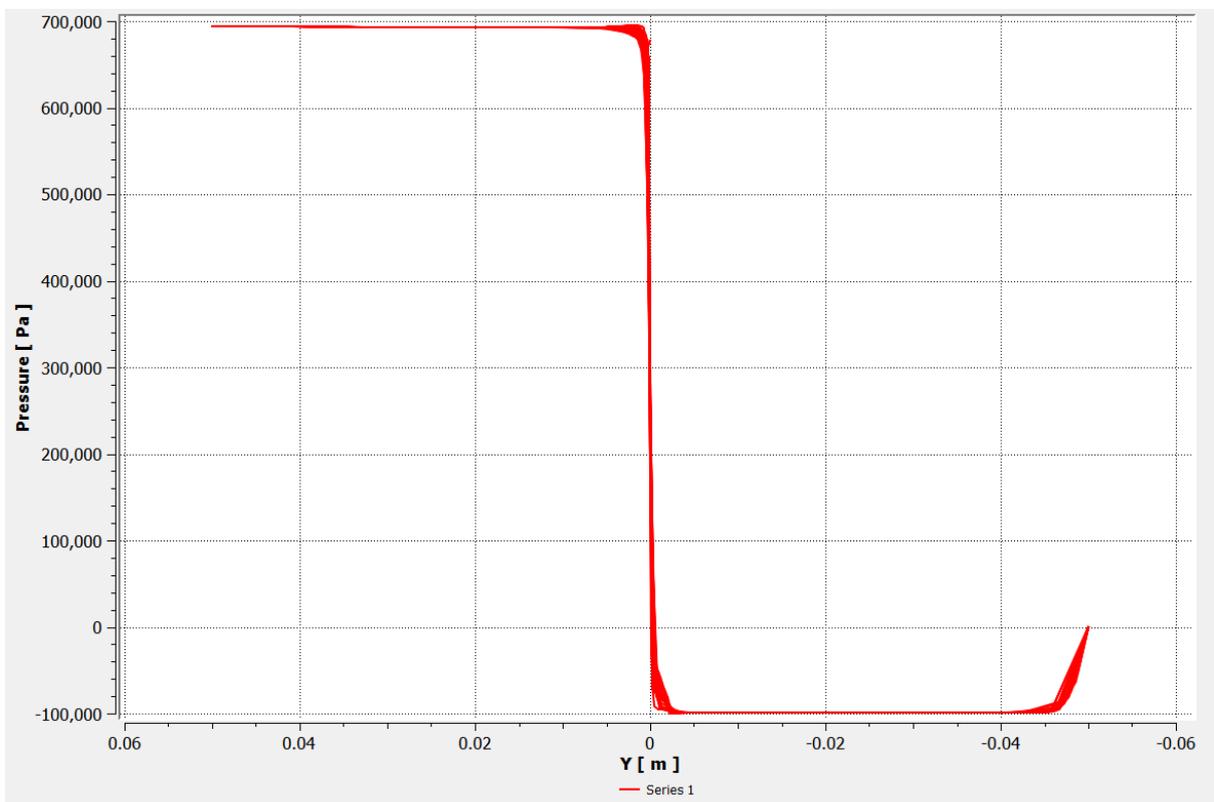

**Figure 22: Pressure Profile - 7 bar**

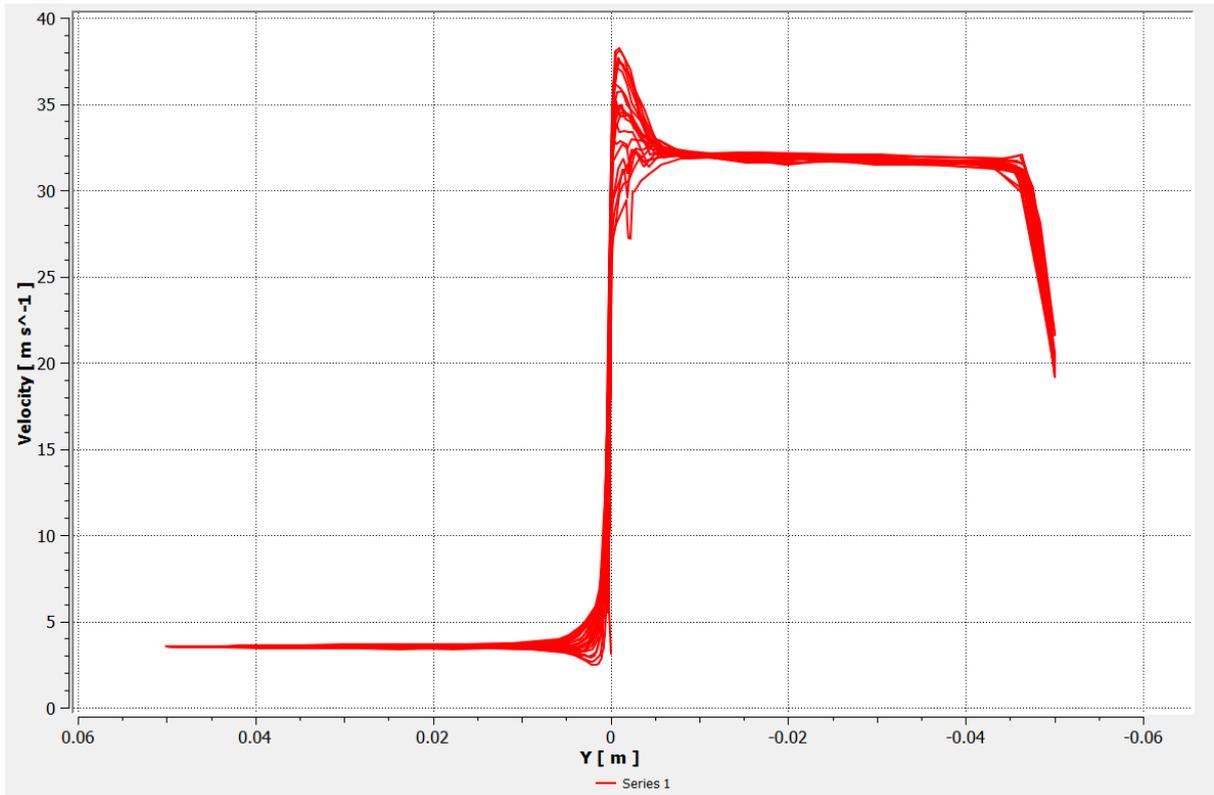

**Figure 23: Velocity Profile - 7 bar**

# 4 bar

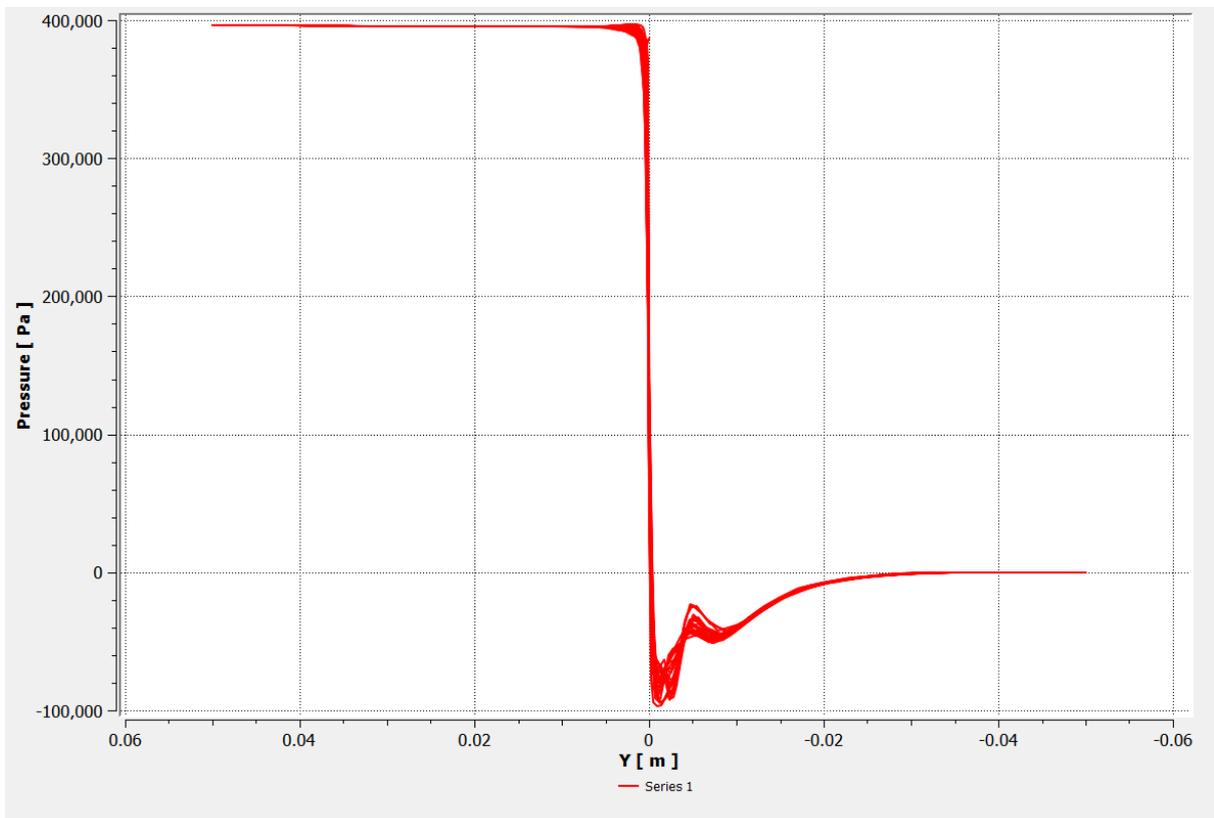

**Figure 24: Pressure Profile - 4 bar**

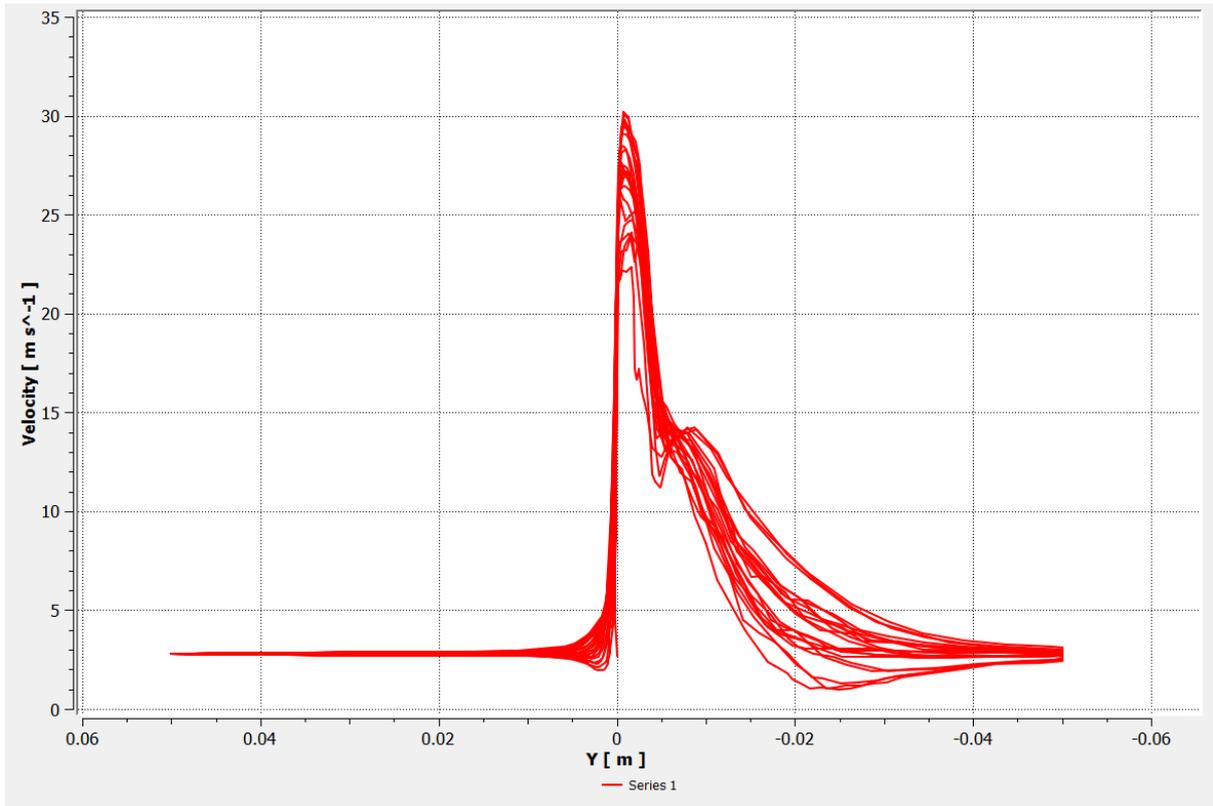

Figure 25: Velocity Profile - 4 bar

# 1 bar

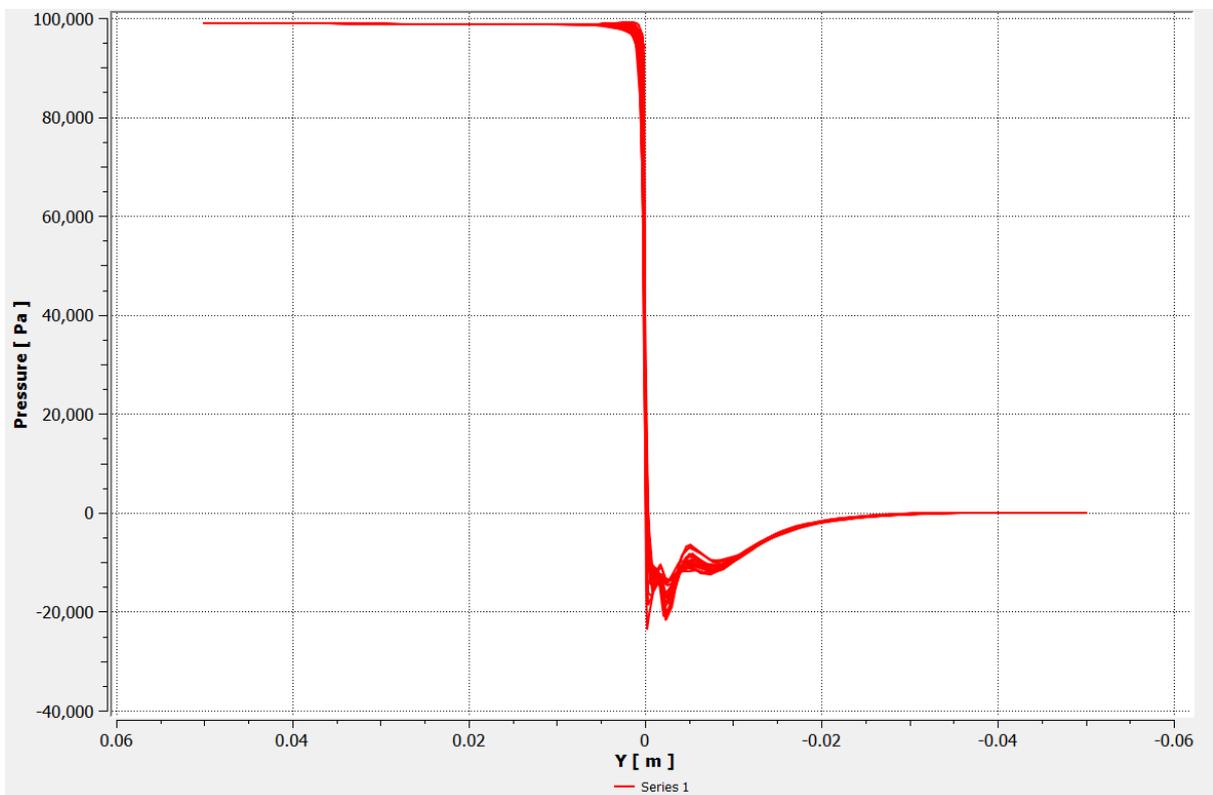

Figure 26: Pressure Profile – 1 bar

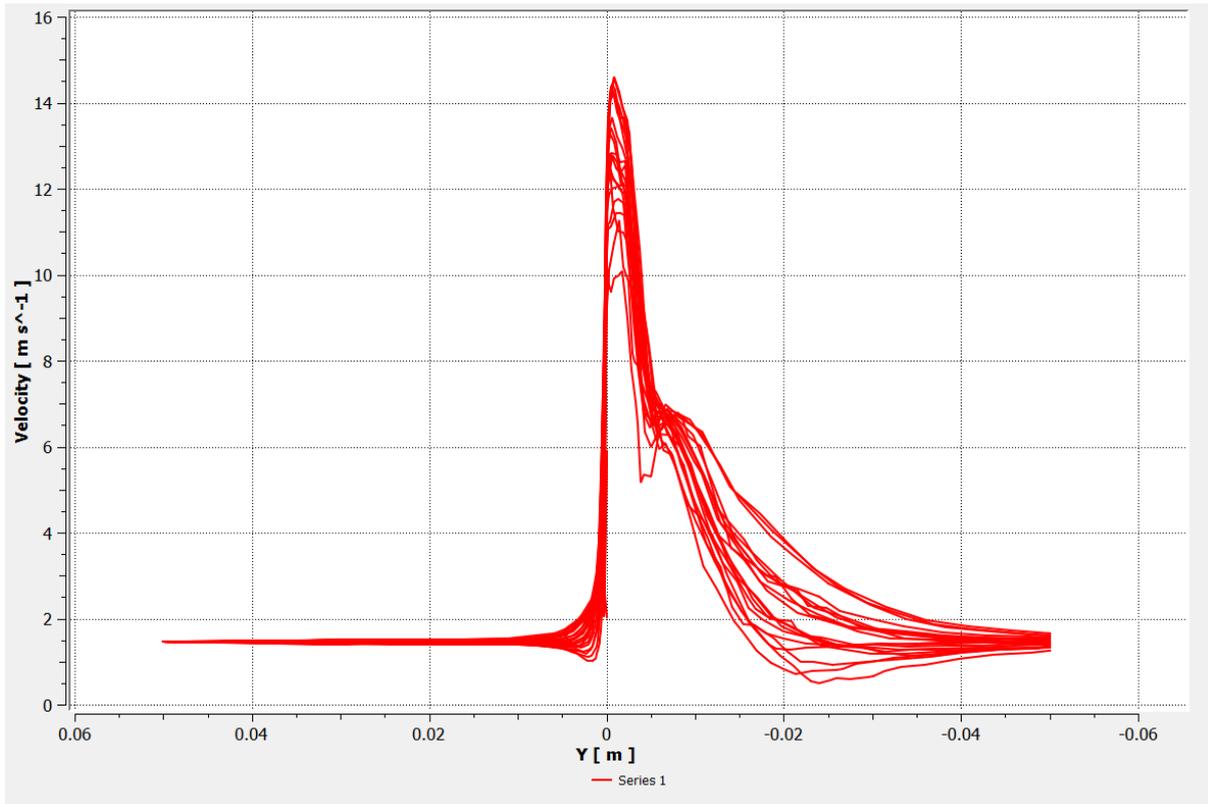

Figure 27: Velocity Profile - 1 bar

# Bubble Radius

| Pressure (bar) | Bubble Radius (m) |
|---:|---|
| 1 | 2.48E-05 |
| 4 | 1.20E-05 |
| 7 | 4.16E-06 |
| 10 | 3.48E-06 |

Table 6: Calculated Bubble Radius for 9 Hole Orifice Plate

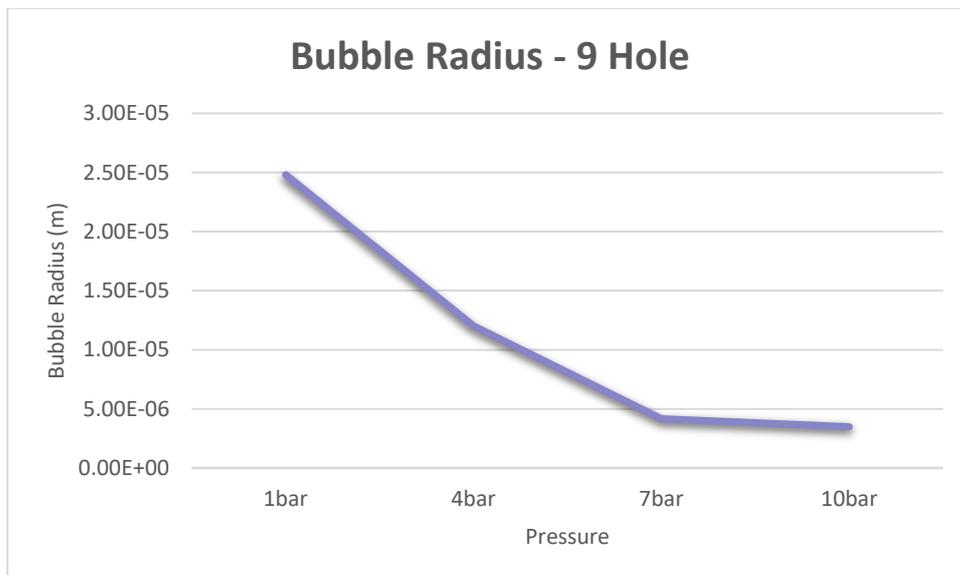

Figure 28: Graphical Representation - Bubble Radius (9 Hole)

## Cavitation Number

| Pressure (bar) | Cavitation Number |
|---:|---|
| 1 | 0.185 |
| 4 | 0.6823 |
| 7 | 0.06266 |
| 10 | 0.06668 |

Table 7: CN - 9 Hole

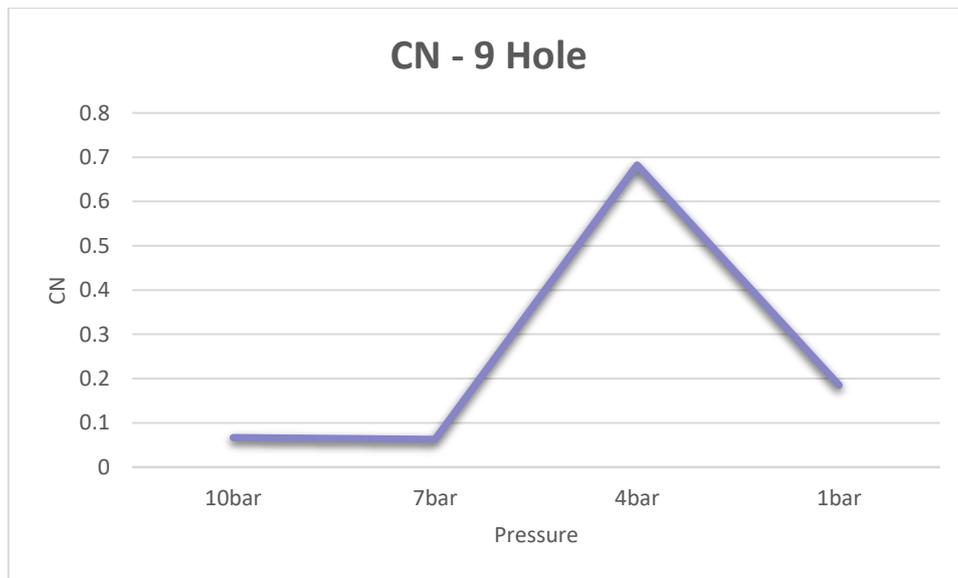

Figure 29: CN - 9 Hole

## 5 Hole

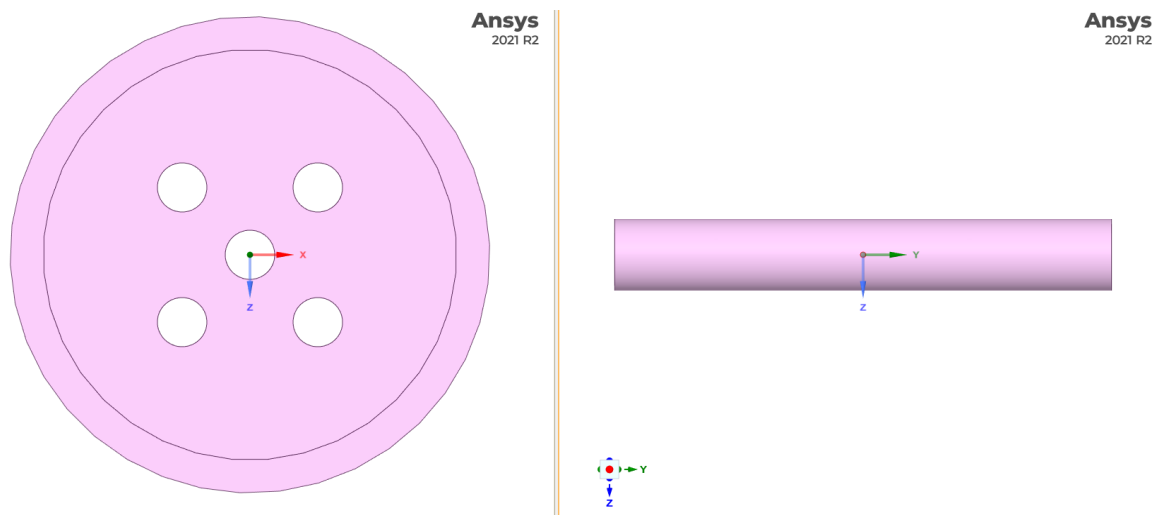

Figure 30: 5 Hole Orifice Geometry

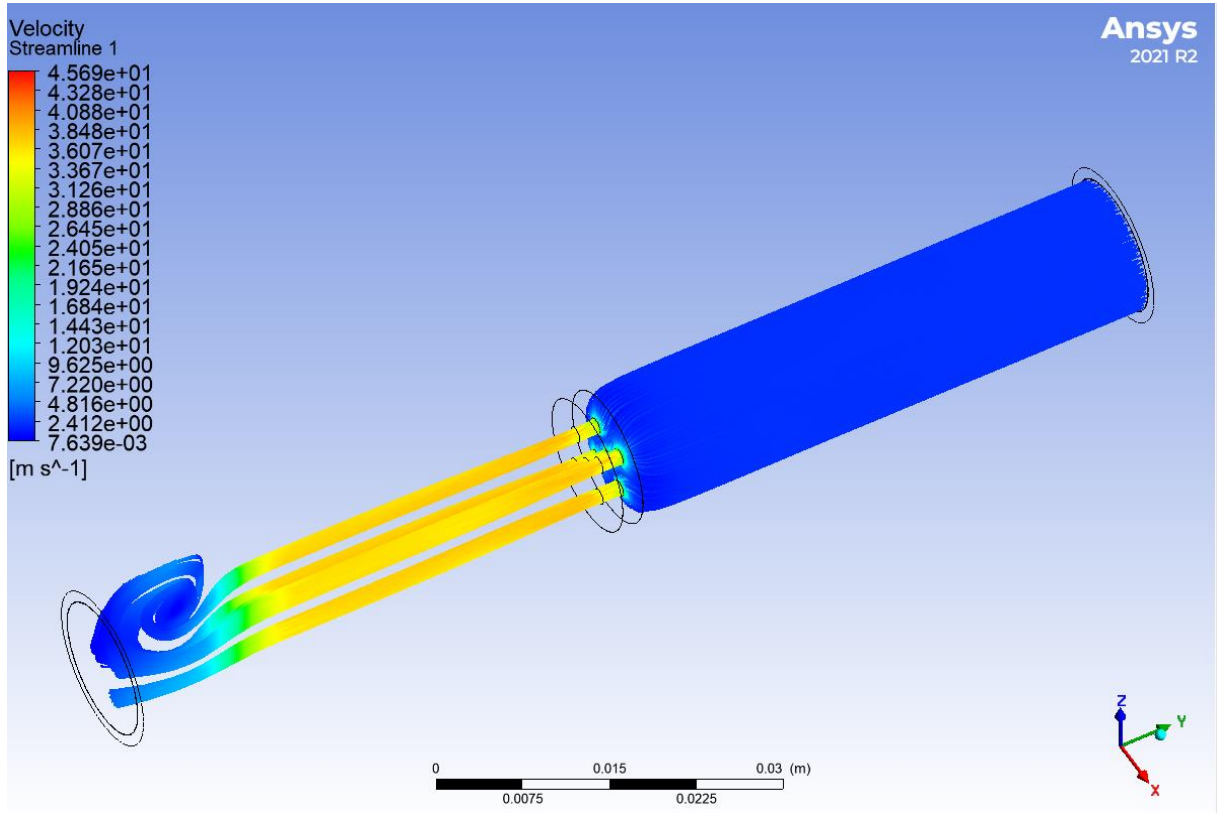

**Figure 31: Velocity Streamline - 5 Hole**

# 10 bar

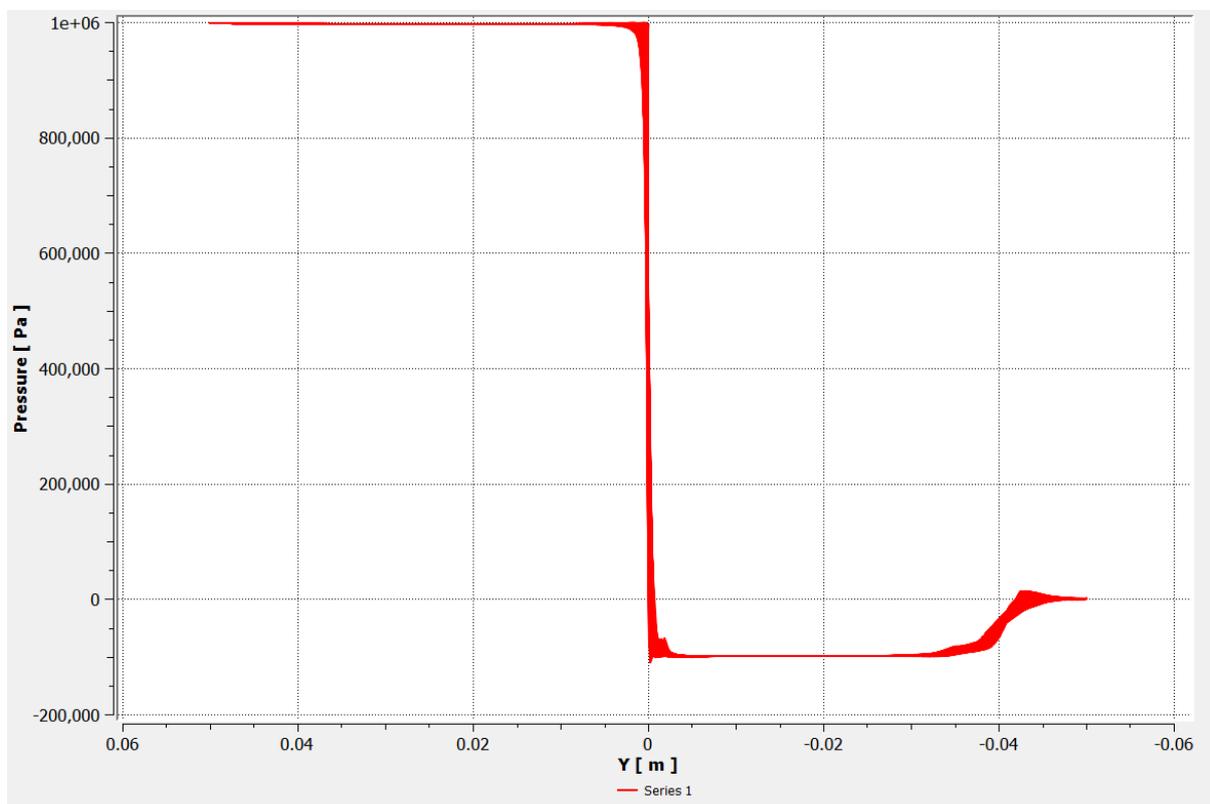

**Figure 32: Pressure Profile - 10 bar**

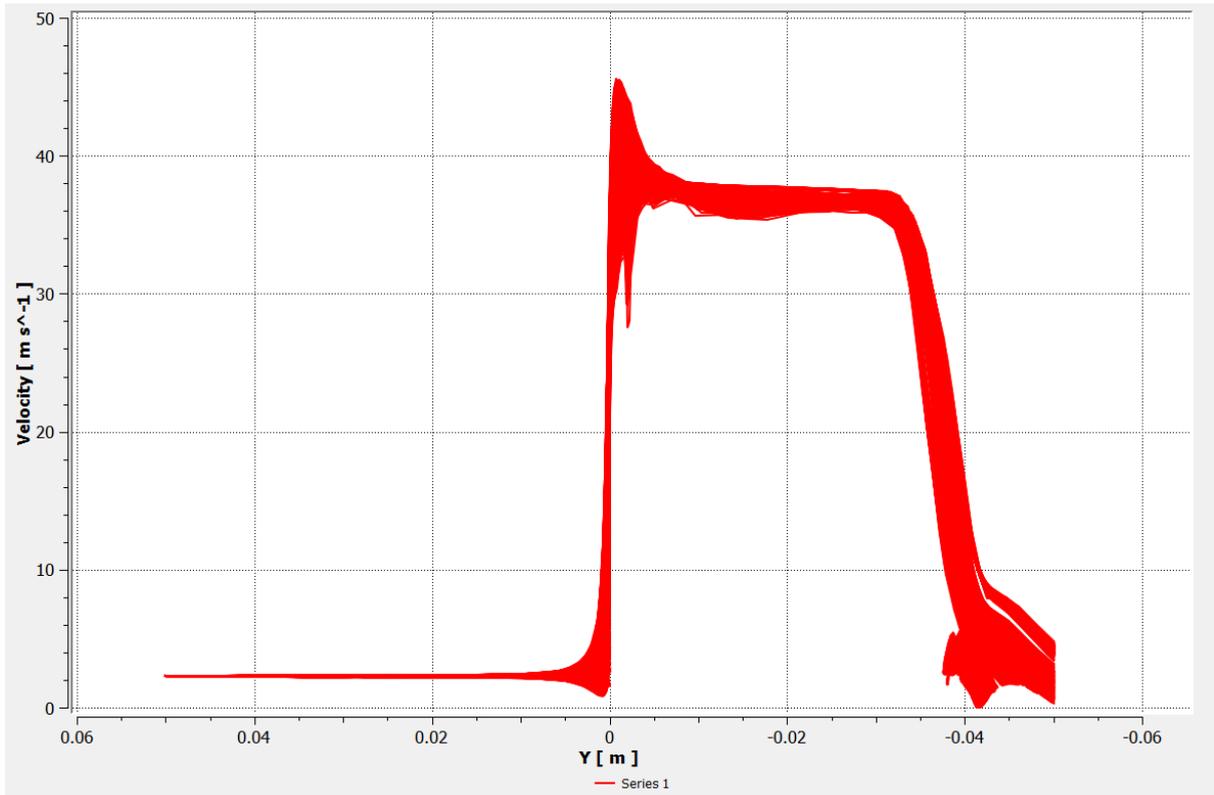

**Figure 33: Velocity Profile - 10 bar**

# 7 bar

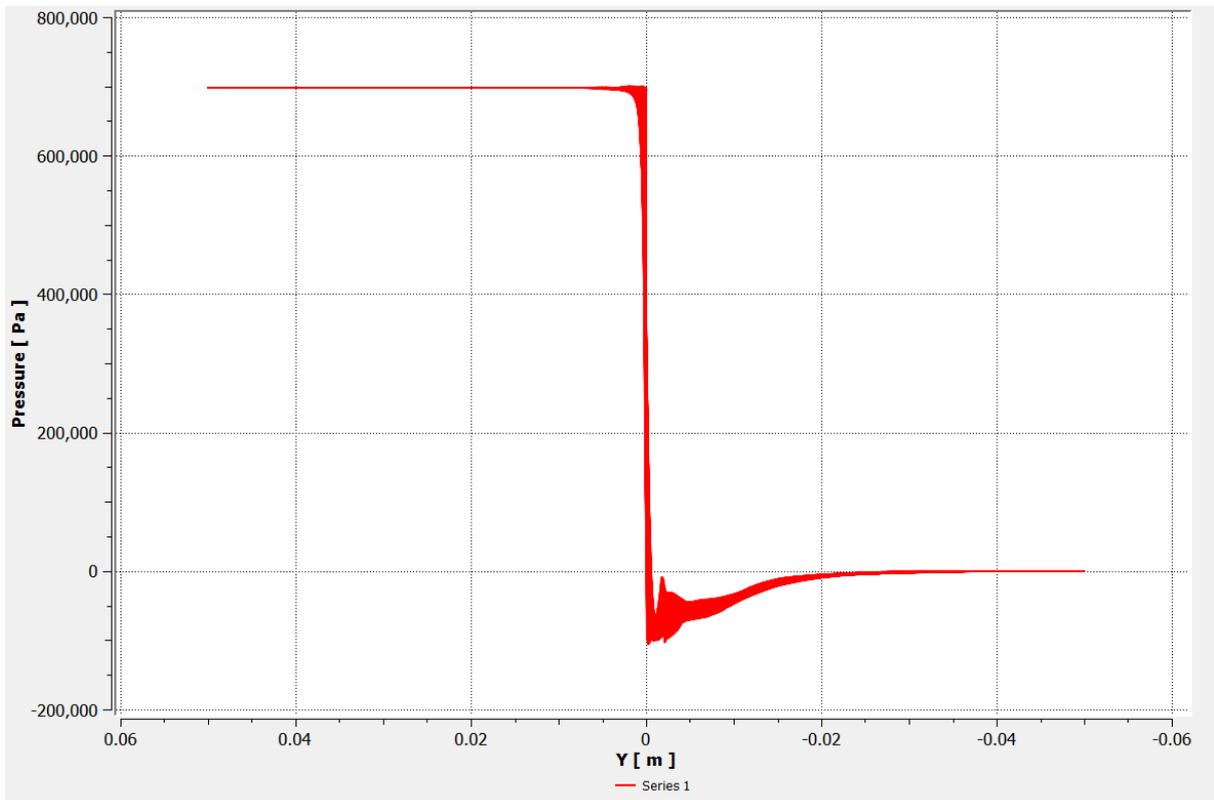

**Figure 34: Pressure Profile - 7 bar**

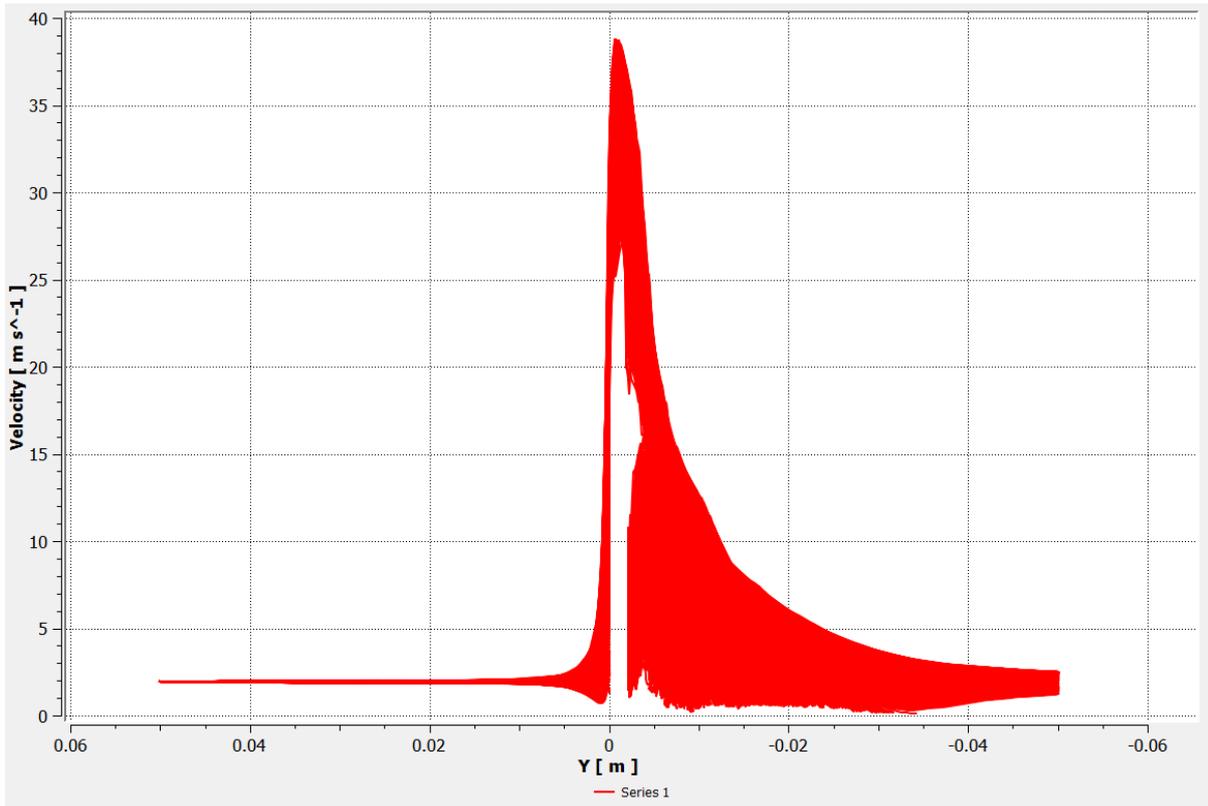

Figure 35: Velocity Profile - 7 bar

# 4 bar

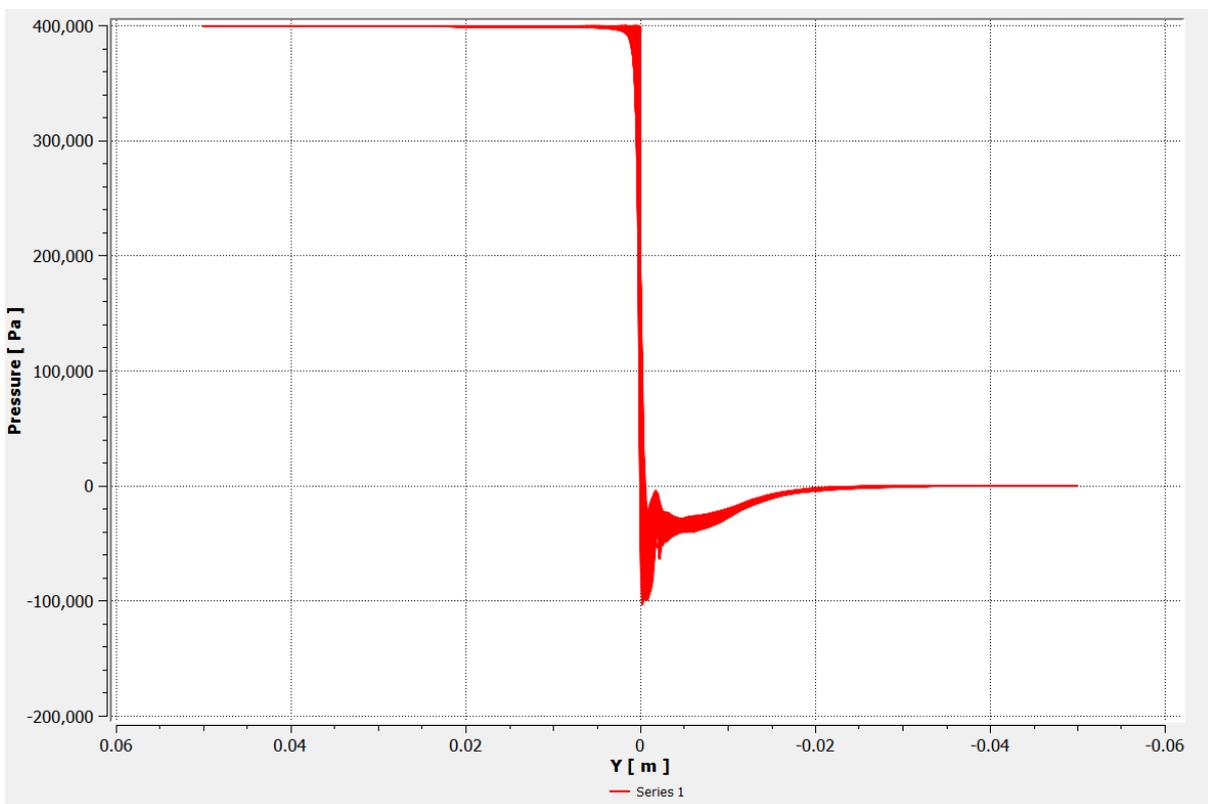

Figure 36: Pressure Profile - 4 bar

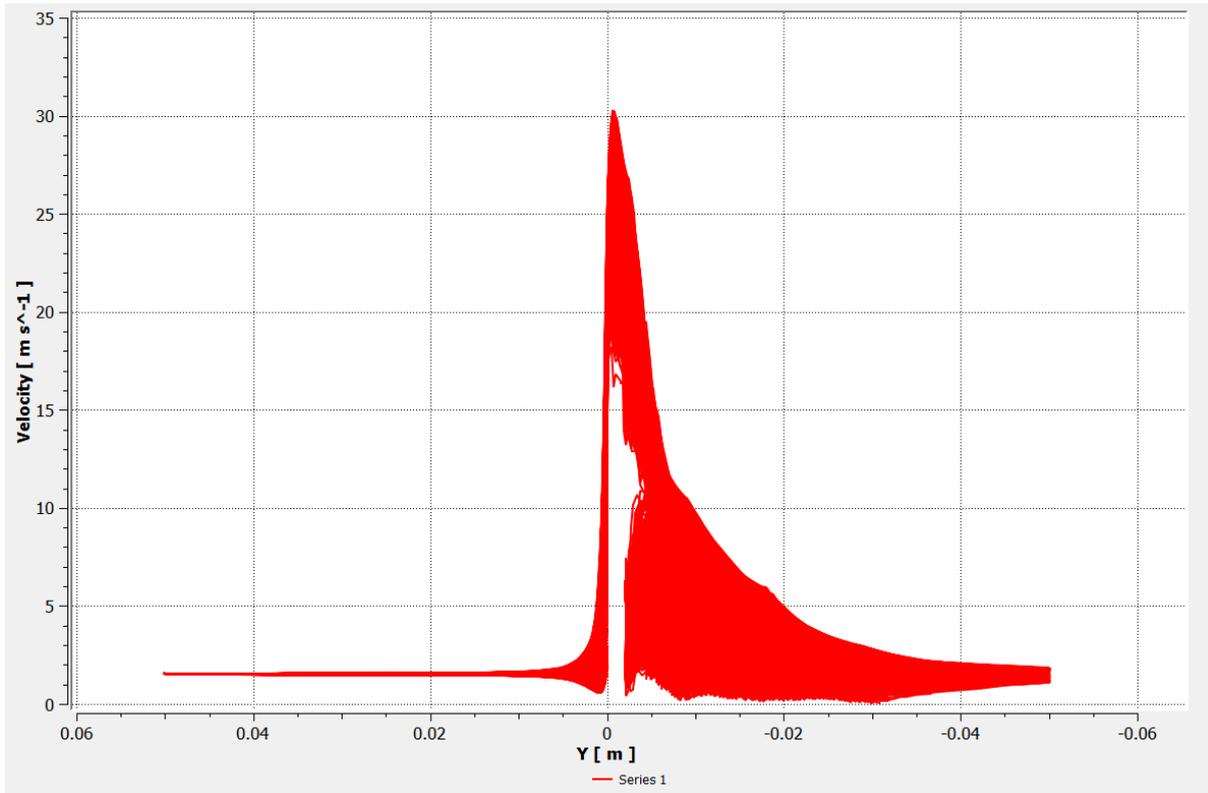

**Figure 37: Velocity Profile - 4 bar**

# 1 bar

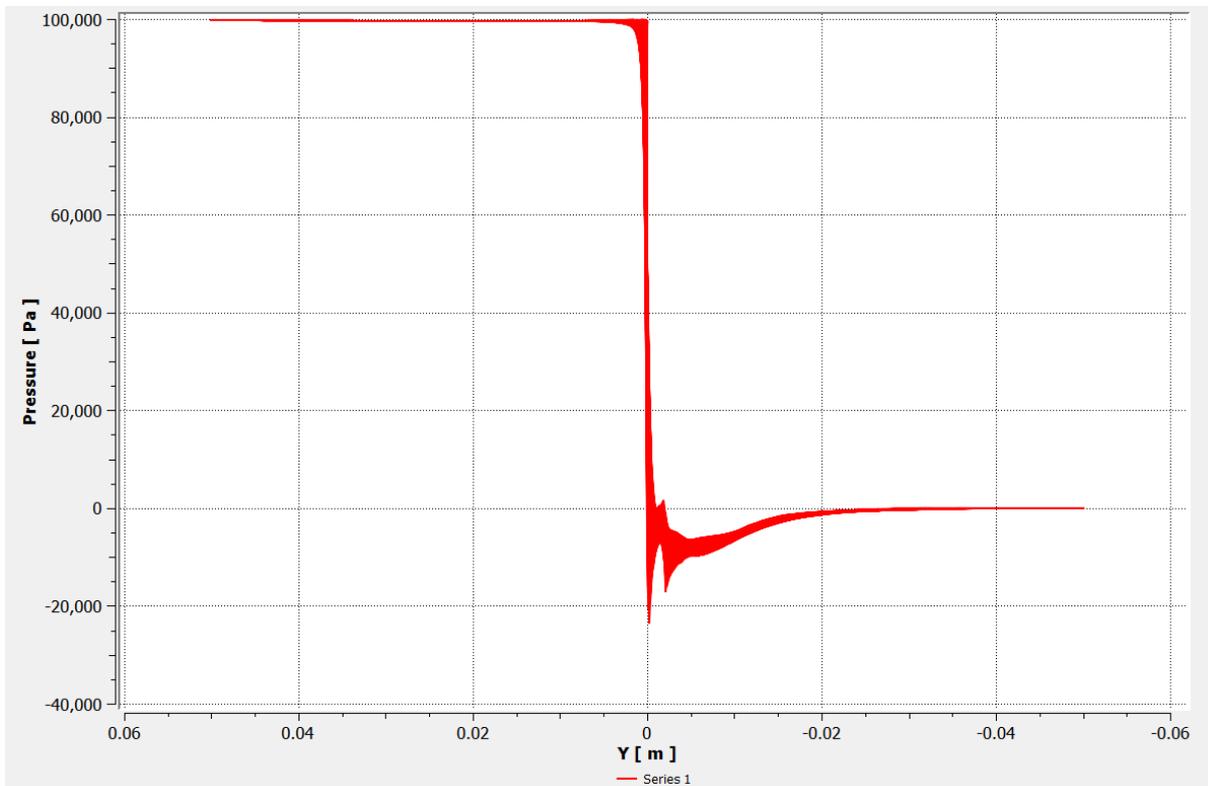

**Figure 38: Pressure Profile - 1 bar**

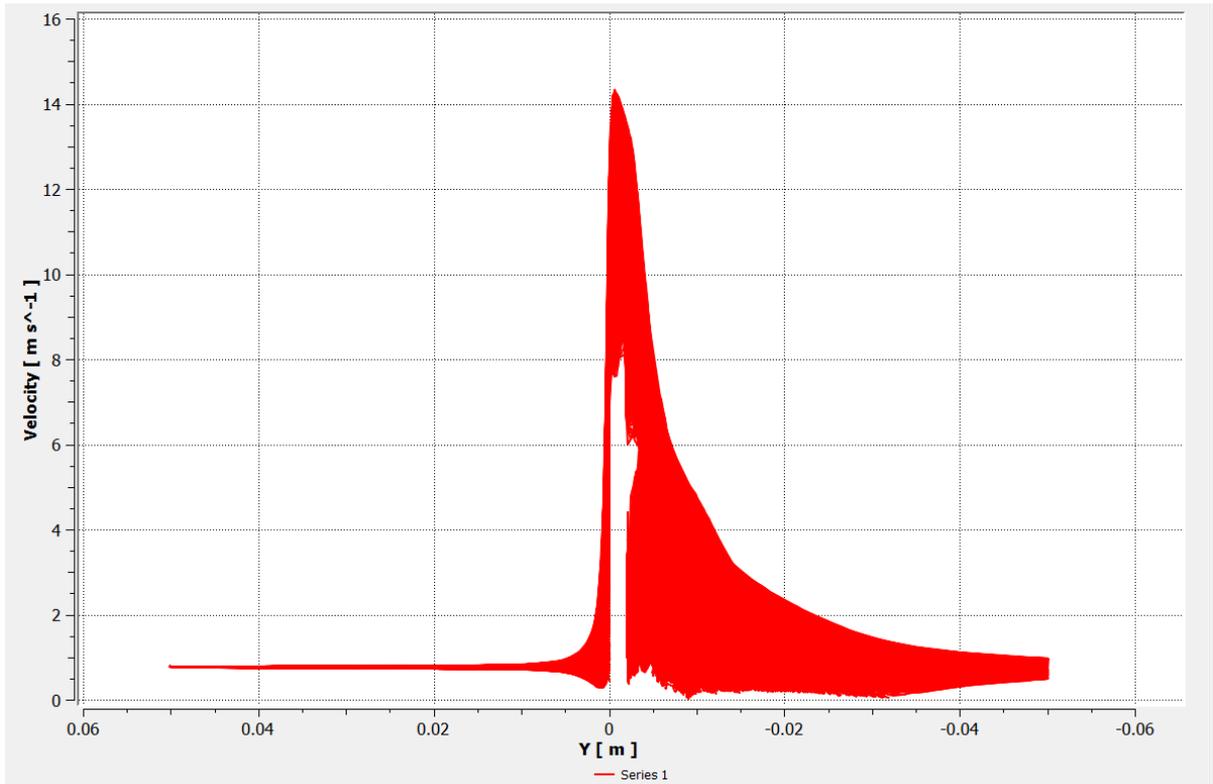

**Figure 39: Velocity Profile - 1 bar**

# Bubble Radius

| Pressure (bar) | Bubble Radius (m) |
|---:|---|
| 1 | 2.54E-05 |
| 4 | 1.22E-05 |
| 7 | 9.46E-06 |
| 10 | 8.03E-06 |

**Table 8: Calculated Bubble Radius for 5 Hole Orifice Plate**

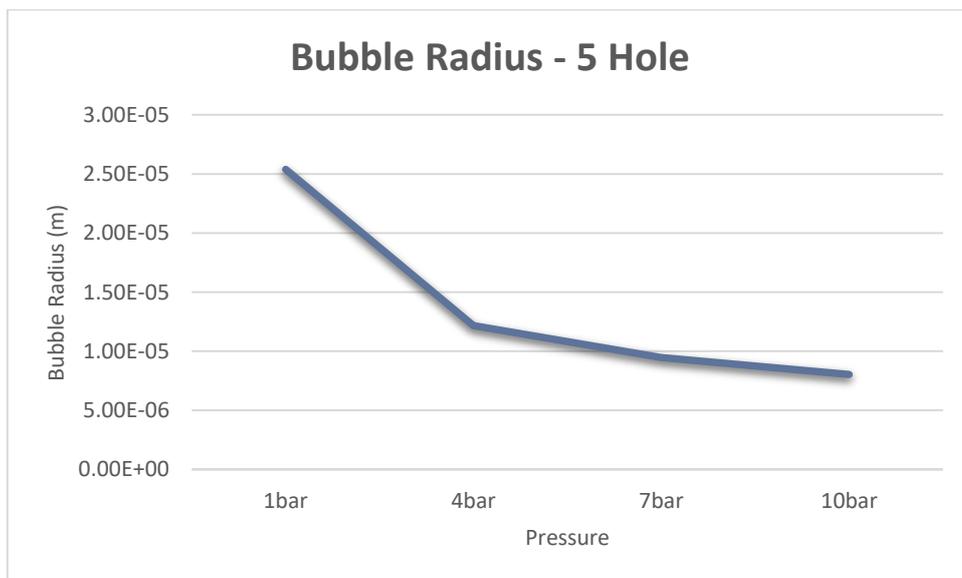

**Figure 40: Graphical Representation - Bubble Radius (5 Hole)**

## Cavitation Number

| Pressure (bar) | Cavitation Number |
|---:|---|
| 1 | 0.8991 |
| 4 | 0.8255 |
| 7 | 0.8034 |
| 10 | 0.2844 |

Table 9: CN - 5 Hole

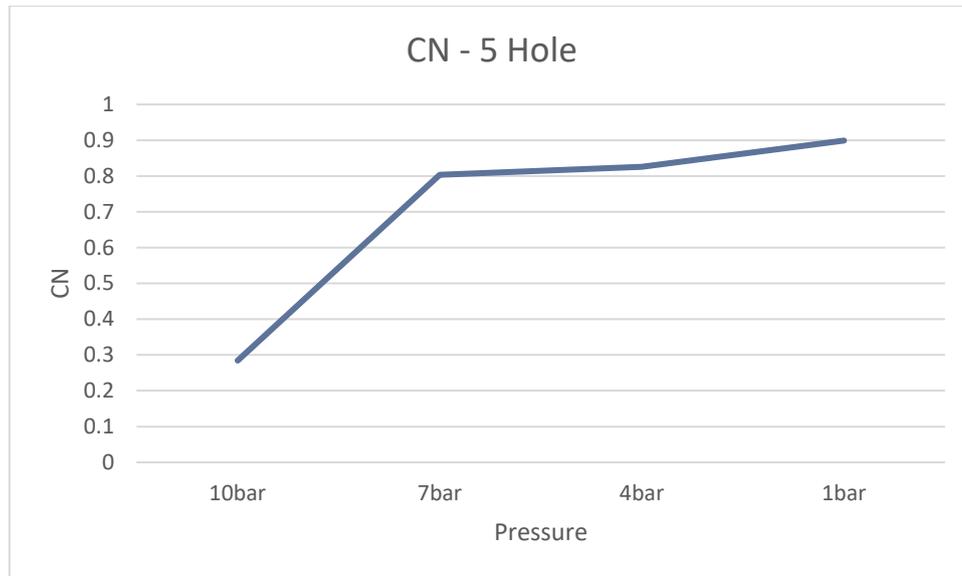

Figure 41: CN - 5 Hole

# Results and Discussion

A total of 3 orifice plates were observed – 10-, 9- and 5-hole. With each orifice plate, 4 pressure profiles were compared – 10, 7, 4 and 1 bar for each Pressure and Velocity Profiles were compared. By the use of Schner-Sauer cavitation model, we were able to observe and calculate the values for Cavitation Number and Bubble Radius. Calculation of Bubble Radius was done via a MATLAB Simulink chart. Similarly, for Cavitation Number, a MATLAB Simulink chart was generated. Calculation of CN involves a variable, recovered pressure which was obtained by Bernoulli' Equation which was further processed into MATLAB to obtain the CN.

Following is the calculated values of all the required parameters.

| | Vel_In (m/s) | Vel_Out (m/s) | Pressure_In (bar) | Pressure_R (bar) | Max Vel (m/s) | CN number | Bubble Radius (m) |
|---|---|---|---|---|---|---|---|
| | | | *10 Holes* | | | | |
| **10bar** | 0.15 | 40 | 10 | 1.9 | 53.52 | 0.1329 | 2.49E-05 |
| **7bar** | 0.108 | 34.42 | 7 | 1.076 | 45.86 | 0.1 | 1.20E-05 |
| *4bar* | 2.2 | 8.5 | 4 | 3.615 | 30.75 | 0.7595 | 8.02E-06 |
| *1bar* | 1.55 | 4.7 | 1 | 0.877 | 14.83 | 0.7748 | 6.81E-06 |
| *9 Holes* | | | | | | | |
|  | Vel_In (m/s) | Vel_Out (m/s) | Pressure_In (bar) | Pressure_R (bar) | Max Vel (m/s) | CN number | Bubble Radius (m) |
| *10bar* | 0.25 | 35.46 | 10 | 3.7 | 105 | 0.06668 | 2.48E-05 |
| *7bar* | 0.207 | 30.11 | 7 | 2.467 | 88.3 | 0.06266 | 1.20E-05 |
| *4bar* | 1.73 | 12.3 | 4 | 3.229 | 30.65 | 0.6823 | 4.16E-06 |
| *1bar* | 1.12 | 4.21 | 1 | 0.2286 | 14.86 | 0.185 | 3.48E-06 |
| *5 Holes* | | | | | | | |
|  | Vel_In (m/s) | Vel_Out (m/s) | Pressure_In (bar) | Pressure_R (bar) | Max Vel (m/s) | CN number | Bubble Radius (m) |
| *10bar* | 1.6 | 37.4 | 10 | 2.993 | 45.69 | 0.2844 | 2.54E-05 |
| *7bar* | 0.9 | 13.3 | 7 | 6.106 | 38.91 | 0.8034 | 1.22E-05 |
| *4bar* | 0.85 | 6.12 | 4 | 3.809 | 30.28 | 0.8255 | 9.46E-06 |
| *1bar* | 0.6 | 2.5 | 1 | 0.967 | 14.48 | 0.8991 | 8.03E-06 |

**Table 10: Calculated Parameters**

# Bubble Radius Comparison

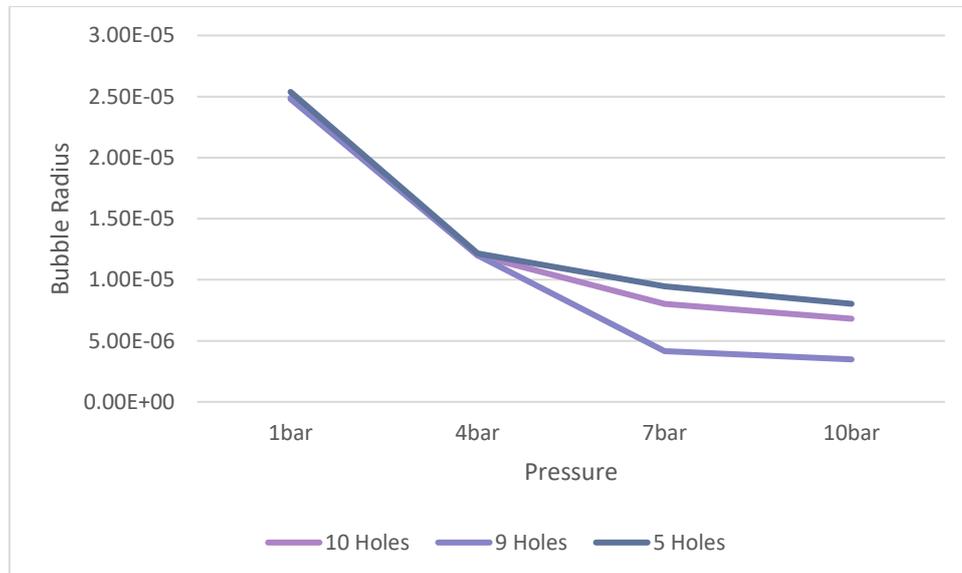

**Figure 42: Bubble Radius Comparison**

## Cavitation Number Comparison

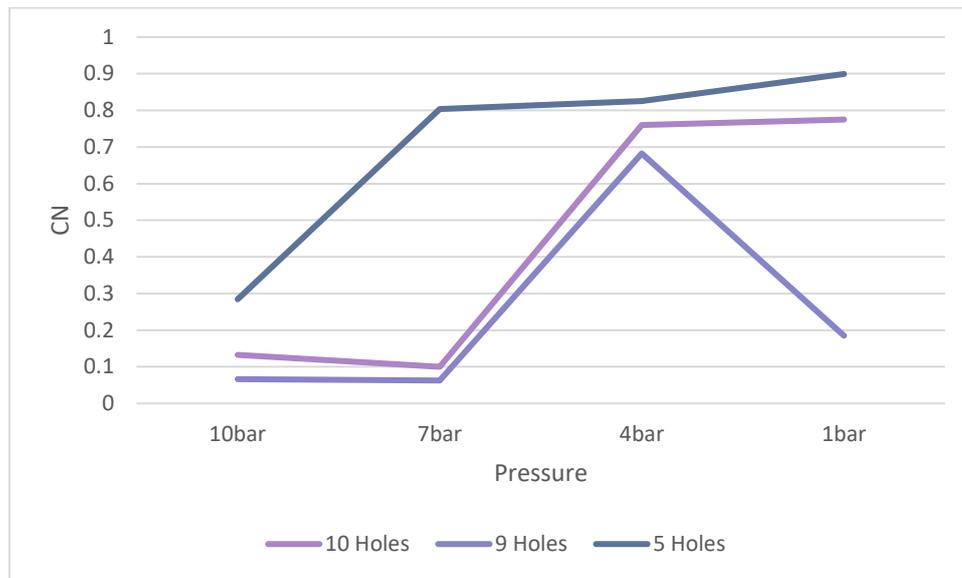

Figure 43: CN Comparison

# Conclusion

A total of 3 orifice plates were observed – 10-, 9- and 5-hole. With each orifice plate, 4 pressure profiles were compared – 10, 7, 4 and 1 bar for each Pressure and Velocity Profiles were compared. Shnerr-Sauer Model was used in the ANSYS Fluent for CFD. An extensive comparison has been made between every model with respect to Pressure and Velocity Profiles, Bubble Radius and Cavitation Number.

Increase in pressure has shown decrease in the size of the Bubble Radius. Hence, for every model, at 10 bar, we got the least value for Bubble Radius. Although, it has been also observed that, at 4 and 1 bar pressure, the bubble radius is almost same which can be observed by the overlapping lines in the comparison graph.

Similarly, increase in pressure has shown decrease in the cavitation number formed. Except for 5-hole orifice plate, at 1 bar pressure, an unprecedented observation is visible which needs to be diagnosed further.

# Bibliography

Babu K J Mahendra, Gowda C J Gangadhara and Ranjith K Numerical Study on Performance Characteristics of Multihole Orifice Plate [Journal]. - Mandya : IOP Conference Series: Materials Science and Engineering, 2018. - Vol. 376.

Bashir Tausif A. [et al.] The CFD Driven Optimisation of a Modified Venturi for Cavitational Activity [Journal]. - Mumbai : The Canadian Journal of Chemical Engineering, 2011. - Vol. 89.

Brennen Christopher E. Cavitation and Bubble Dynamics [Book]. - New York : Oxford University Press, 1995.

Dabiri S., Sirignano W. A. and Joseph D. D. Cavitation in an orifice flow [Journal]. - Minneapolis : Physics of Fluid , 2007. - Vol. 19.

Ebrahimi Behrouz [et al.] Characterization of high-pressure cavitating flow through a thick orifice plate in a pipe of constant cross section [Journal]. - Houston : Elsevier Masson SAS, 2017. - Vol. 114.

Jin Zhi-jiang [et al.] Cavitating Flow through a Micro-Orifice [Journal]. - Hangzhou : Micromachines , 2019. - Vol. 10.

Jithish KS and Kumar PV Ajay Analysis of turbulent flow through an orifice meter using experimental and computational fluid dynamics simulation approach - A case study [Journal]. - Kadayiruppu : International Journal of Mechanical Engineering Education, 2015. - Vol. 0.

Kuldeep and Saharan Virendra Kumar Computational study of different venturi and orifice type hydrodynamic cavitating devices [Journal]. - Jaipur : Elsevier , 2016. - Vol. 28.

Lindau Jules W. [et al.] High Reynolds Number, Unsteady, Multiphase CFD Modeling of Cavitating Flows [Journal]. - University Park : Journal of Fluids Engineering, 2002. - Vol. 124.

S.Shah Manish, Jyeshtharaj B. Joshi Avtar S. Kalsi, C.S.R.Prasad and Shukla Daya S. Analysis of flow through an orifice meter: CFD simulation [Journal]. - Mumbai : Elsevier Ltd, 2011. - Vol. 71.

Saharan Virendra Kumar [et al.] Effect of geometry of hydrodynamically cavitating device on degradation of orange-G [Journal]. - Mumbai : Elsevier, 2012. - Vol. 20.

Simpson Alister and Ranade Vivek V. Modelling of Hydrodynamic Cavitation with Orifice: Influence of different orifice designs [Journal]. - Belfast : Chemical Engineering and Research Design , 2018.

Venturi Meter Vs Orifice Meter: What Is The Difference? [Online] // The Engineer's Perspective . - May 2022. - https://www.theengineersperspectives.com/venturi-vs-orifice-what-is-the-difference/.

Yan Y. and Thorpet R. B. Flow Regime Transitions due to Cavitation in the Flow through an Orifice [Journal]. - Cambridge : International Journal Multiphase Flow, 1990. - Vol. 16.

Zhang Yu [et al.] Cavitation optimization of single-orifice plate using CFD method and neighborhood cultivation genetic algorithm [Journal]. - Chengdu : Elsevier Korea LLC, 2021. - Vol. 54.